\documentclass[en,src,std]{monograph}

\usepackage{cite}
\usepackage[T1]{fontenc}
\usepackage{url}

\usepackage{graphicx} 
\usepackage{graphics} 
\usepackage{xcolor}   
\usepackage{rotating}
\usepackage[colorlinks,hyperindex,plainpages=false]{hyperref}
\usepackage{array,enumerate}
\usepackage{amsmath}
\usepackage{bbm}
\usepackage{stmaryrd}

\usepackage{tabularx}
\usepackage{amsmath,amsthm,amsfonts,amssymb,amsxtra}
\let\mathbb\undefined  
\usepackage{bbold}     

\usepackage{realboxes}

\newcommand{\theoryfont}{\normalfont\selectfont}

\normalfont

\usepackage[titles]{tocloft}

\DeclareFontFamily{OMX}{MnSymbolE}{}
\DeclareSymbolFont{largesymbolsMn}  {OMX}{MnSymbolE}{m}{n}
\SetSymbolFont{largesymbolsMn}{bold}{OMX}{MnSymbolE}{b}{n}

\DeclareFontShape{OMX}{MnSymbolE}{m}{n}{
    <-6>  MnSymbolE5
   <6-7>  MnSymbolE6
   <7-8>  MnSymbolE7
   <8-9>  MnSymbolE8
   <9-10> MnSymbolE9
  <10-12> MnSymbolE10
  <12->   MnSymbolE12}{}
\DeclareFontShape{OMX}{MnSymbolE}{b}{n}{
    <-6>  MnSymbolE-Bold5
   <6-7>  MnSymbolE-Bold6
   <7-8>  MnSymbolE-Bold7
   <8-9>  MnSymbolE-Bold8
   <9-10> MnSymbolE-Bold9
  <10-12> MnSymbolE-Bold10
  <12->   MnSymbolE-Bold12}{}
\DeclareMathDelimiter{\lgroup}{\mathopen}{largesymbolsMn}{'242}{largesymbolsMn}{'242}
\DeclareMathDelimiter{\rgroup}{\mathopen}{largesymbolsMn}{'243}{largesymbols}{'243}

\DeclareMathAlphabet{\mathpzc}{OT1}{pzc}{m}{it}

\newcounter{definition}[section]
\renewcommand\thedefinition {\arabic{section}.\arabic{definition}}

\newskip\thmheadsep
\newenvironment{edefinition}%
{\trivlist \setlength{\topsep}{0pt} \setlength{\partopsep}{0mm}\item[]%
\refstepcounter{definition}\refstepcounter{equation}%
\begin{minipage}{\linewidth}
       \hbox to \linewidth\bgroup \hbox to 100pt{\scshape Definition \thedefinition.\hfil} $
         \displaystyle\hfil}%
        {$\hfil 
         \displaywidth\linewidth\hbox to 100pt{\hfil\@eqnnum}%
       \egroup%
     \end{minipage}\endtrivlist}

{\par\vspace{-5pt}\noindent%
\refstepcounter{definition}\refstepcounter{equation}%
\begin{minipage}{\linewidth}

\begin{tabular*}{\textwidth}{p{100pt} p{\textwidth-200pt} p{100pt}}
\hbox to 100pt{\scshape Definition \thedefinition.\hfil} &%
\hfil$\displaystyle %
}%
{%
$ \hfil & %
 \hbox to 100pt{\hfil\@eqnnum}%
\end{tabular*}

     \end{minipage}\\[10pt]}

\newdimen\csize
\newdimen\cindent
\newsavebox\eqbox
\newlength\eqlen
\newsavebox\cbox
\newlength\clen
\newenvironment{cequation}[1]{\ignorespaces
\noindent\vskip\abovedisplayskip %
\refstepcounter{equation}%
\hbox to \linewidth\bgroup %
\hbox to \csize{\hfil$\displaystyle\forall\;#1\relax:$\hskip \cindent} %
\Savebox\eqbox\bgroup$\displaystyle}{$\egroup\ignorespaces
\settowidth\eqlen{\usebox\eqbox}
\ifdim\eqlen<\clen{%
\hfil\usebox\eqbox\hfil
\displaywidth\linewidth\hbox to \csize{\hfil\@eqnnum}%
}\else{%
\usebox\eqbox\hfil
\displaywidth\linewidth\hbox to 10mm{\hfil\@eqnnum}%
}%
\fi
\egroup %
\vskip \belowdisplayskip}

{%
$\hfil \displaywidth\linewidth\hbox to \csize{\hfil\@eqnnum}%
\egroup %
\vskip \belowdisplayskip 
}

\newcounter{test}[section]

\renewcommand{\thetest} {\arabic{section}.\arabic{test}}

\newskip\thmheadsep
\newcounter{statement}
\renewcommand\thestatement {\arabic{section}.\arabic{statement}}

\newskip\thmheadsep
\newenvironment{statement}{%
\trivlist  \setlength{\topsep}{15pt} \item
\refstepcounter{statement}
\begin{minipage}{\linewidth}
\mbox{\scshape Statement \thestatement.\hspace{4pt}}  \theoryfont
}{%
\end{minipage}\endtrivlist
}

\newtheoremstyle{itps}%
   {20pt}{15pt}
   {\theoryfont \itshape}
   {0pt}
   {\scshape}{.}
   {1ex plus 0pt minus .2ex}
   {}

\theoremstyle{itps}

\newtheorem{interpretation}{Interpretation}[section]

\newtheoremstyle{form}%
   {15pt}{20pt}
   {\theoryfont\itshape}
   {0pt}
   {\scshape}{.}
   {1ex plus 0pt minus .2ex}
   {}

\theoremstyle{form}

\newtheorem{formalism}{Formalism}

\newcommand{\twocolumneq}[2]{
\noindent\begin{subequations}
\newline\parbox{\linewidth}%
{%
\parbox{0.5\linewidth}%
{#1}%
\parbox{0.5\linewidth}%
{#2}%
}%
\end{subequations}
}

\newcommand{\parbreak}{\hfill\nobreak$\infty$\nobreak\vskip 17pt}
\newcommand{\qparbreak}{\vspace{-\baselineskip}\nobreak\hfill\nobreak$\infty$\nobreak\vskip 17pt}

\newcommand{\dsp}{\quad|\quad}
\newcommand{\csp}{\;\;|\;\;}
\newcommand{\bsp}{\;|\;}
\newcommand{\asp}{\,\raisebox{1pt}{\tiny|}\,}

\renewcommand{\asp}{,}

\newcommand{\tand}{\text{ and }}
\newcommand{\tor}{\text{ or }}

\renewcommand{\iff}{\;\Longleftrightarrow\;}
\newcommand{\then}{\;\Longrightarrow\;}

\newcommand{\eqspace}{\\[5pt]}

\let\set\undefined

\newcommand{\eqd}{\,\stackrel{\text{\tiny def}}{=}\,}

\newcommand{\rld}{\stackrel{\text{\tiny def}}{\iff}}

\newcommand{\cpt}{\rightleftharpoons}

\newcommand{\ctrd}{\mathrel{\shortrightarrow\kern-1pt \shortleftarrow}}

\newcommand{\st}{\dag}

\newcommand{\s}{^{\st}{}}

\newcommand{\set}[1]{\ensuremath{\mathbbmss{#1}}}

\newcommand{\sdef}[1]{\ensuremath{\left\{#1\right\}}}

\newcommand{\prp}[1]{\ensuremath{\widehat{#1}}}

\newcommand{\stp}[1]{\ensuremath{\left[#1 \right]}}

\newcommand{\bra}[1]{\ensuremath{\kern-1.2pt \mathinner{\langle#1|}\kern-1pt}}
\newcommand{\ket}[1]{\ensuremath{\kern-1pt\mathinner{|#1\rangle}\kern-1.2pt }}

\def\stp#1{\ensuremath{\kern-1pt\mathinner{[\kern 0.3pt#1\kern 0.2pt]}\kern-1pt}}

\newcommand{\braket}[1]{\ensuremath{\kern-1.2pt\mathinner{\langle{#1}\rangle}\kern-1.2pt}}

\newcommand{\Braket}[1]{\ensuremath{\kern-1.2pt\mathinner{\left\langle #1\right\rangle}\kern-1.2pt}}

\newcommand{\unity}[0]{\ensuremath{\mathbf{\hat{1}}}}
\newcommand{\db}[0]{ \lower-.3ex\hbox{$\scriptstyle\mid$}\kern-.1ex}

\newcommand{\rlty}{\stp{\psi}}

\newcommand{\bools}{\set{B}}

\newcommand{\prps}{\set{P}}

\newcommand{\trus}{\set{T}}
\newcommand{\falss}{\set{F}}
\newcommand{\Ntrus}{\ensuremath{\set{T}_{\kern-2pt\mathpzc{N}}}}
\newcommand{\Nfalss}{\ensuremath{\set{F}_{\kern-2pt\mathpzc{N}}}}

\newcommand{\Ptrus}{\ensuremath{\trus_{\kern-2pt\mathpzc{P}}}}
\newcommand{\Pfalss}{\ensuremath{\falss_{\kern-2pt\mathpzc{P}}}}

\newcommand{\Cprps}{\ensuremath{\prps_{\kern-2pt\mathpzc{C}}}}

\newcommand{\Ktrus}[1]{\ensuremath{\set{T}_{\kern-2pt\mathpzc{K}^#1}}}

\newcommand{\Kfalss}[1]{\ensuremath{\set{F}_{\kern-2pt\mathpzc{K}^#1}}}

\newcommand{\ctrs}{\ensuremath{\mathbb{0}}}
\newcommand{\tauts}{\set{1}}

\newcommand{\soks}{\set{K}}

\newcommand{\soas}{\set{S}}

\newcommand{\ets}{\set{U}}

\newcommand{\ops}{\ensuremath{\mathcal{L}}}

\newcommand{\rops}{\ensuremath{\mathcal{R}}}

\newcommand{\pops}{\ensuremath{\rops_{\geq 0}}}

\newcommand{\bras}[0]{\ensuremath{\mathcal{B}}}
\newcommand{\kets}[0]{\ensuremath{\mathcal{K}}}
\newcommand{\pjs}[0]{\ensuremath{\mathcal{P}}}

\newcommand{\Sbras}[0]{\ensuremath{\bras_S}}
\newcommand{\Skets}[0]{\ensuremath{\kets_S}}
\newcommand{\Spjs}[0]{\ensuremath{\pjs_S}}

\newcommand{\rst}[2]{\ensuremath{#1_{\db #2}}}

\newcommand{\oprst}[1]{\rst{\ops}{#1}}

\newcommand{\roprst}[1]{\rst{\rops}{#1}}
\newcommand{\poprst}[1]{\rst{\pops{}}{#1}}
\newcommand{\pjrst}[1]{\rst{\pjs}{#1}}
\newcommand{\ketrst}[1]{\rst{\kets}{#1}}
\newcommand{\brarst}[1]{\rst{\bras}{#1}}
\newcommand{\Sketrst}[1]{\rst{\Skets{}}{#1}}
\newcommand{\Sbrarst}[1]{\rst{\Sbras{}}{#1}}

\newcommand{\poprt}[0]{\rst{\pops{}}{\unity}}

\newcommand{\pjsrt}[0]{\rst{\pjs}{\unity}}

\newcommand{\Spjsrst}[1]{\rst{\Spjs{}}{#1}}

\newcommand{\un}[0]{\ensuremath{\cup}}
\newcommand{\its}[0]{\ensuremath{\cap}}

\newcommand{\tuple}[1]{\ensuremath{\langle#1\rangle}}
\newcommand{\ensSet}[1]{\{\kern-3.7pt \{ #1 \} \kern-3.7pt \}}
\newcommand{\ensTuple}[1]{\langle \kern-2.5pt  \langle #1 \rangle \kern-2.5pt \rangle}

\def\enditemeq{\hfill \hbox{\@eqnnum}$ \par}

\def\@eqnnum{{\normalfont \normalcolor (\theequation)}}

\newcommand{\pand}[0]{\ensuremath{\wedge}}
\newcommand{\por}[0]{\ensuremath{\vee}}

\newcommand{\pnot}[0]{\ensuremath{\neg}}

\newcommand{\pcmp}[1]{\ensuremath{\mathop{\neg}^{#1}}}

\newcommand{\pgeq}[0]{\ensuremath{\sqsupseteq}}
\newcommand{\pleq}[0]{\ensuremath{\sqsubseteq}}

\newcommand{\npgeq}[0]{\ensuremath{\not\sqsupseteq}}

\newcommand{\idsc}{\Subset}

\newcommand{\state}[0]{\ensuremath{\rho}}

\newcommand{\pbs}[0]{\ensuremath{\mathbf{P}}}

\newcommand{\pbr}[1]{\pb{#1}{\state}}

\newcommand{\pb}[2]{\ensuremath{\pbs_{#2} ( #1  ) }}

\newcommand{\pbn}[1]{\pb{#1}{\mathcal{N}}}

\newcommand{\pbk}[2]{\pb{#1}{\mathcal{K}^#2}}

\newcommand{\pbki}[1]{\pbk{#1}{i}}

\newcommand{\pbt}[1]{\pb{#1}{\mathcal{T}}}

\newcommand{\pbns}{\ensuremath{\pbs_{\kern-2pt\mathpzc{N}}}}

\newcommand{\pbks}[1]{\ensuremath{\pbs_{\kern-2pt\mathpzc{K}_#1}}}

\newcommand{\pbkis}{\ensuremath{\pbs_{\kern-2pt\mathpzc{K}^i}}}

\newcommand{\pbts}{\ensuremath{\pbs_{\kern-2pt\mathpzc{T}}}}

\newcommand{\msr}[1]{\ensuremath{\mu  ( #1 ) }}

\newcommand{\norm}[1]{\ensuremath{\| #1 \|}}

\newcommand{\Nset}[0]{\set{N}}

\newcommand{\Rset}[0]{\set{R}}

\newcommand{\Cset}[0]{\set{C}}

\newcommand{\Zset}[0]{\set{Z}}

\newcommand{\Qset}[0]{\set{Q}}

\newcommand{\Xset}[0]{\set{X}}

\newcommand{\Aset}[0]{\set{A}}

\newcommand{\Bset}[0]{\set{B}}

\title{On The Foundations of the\\ Final Theory of Physics}

\date{11 de junho de 2012}

\author{Frederico R. Pfrimer}

\begin{document}
\frontmatter
\maketitle
\tableofcontents





\mainmatter

\chapter{Introduction}

Foundations have always been a problem. You may just not have noticed or been involved in these issues, even so, it is a fact. To the astonishment of some, this is a fact that permeates all main areas of human knowledge, including Physics, Computer Science, Philosophy and even Mathematics. However, applications follow exactly the opposite direction of foundations, and that distance us from these issues and allows the progress of science and society occur before we get answers to the most fundamental questions. It is like this, and so should be. It doesn't mean that we will never find the answers, but that we must persist in the pursuit,  and, at the same time, proceed in their absence, without, however, forgetting that the question remains unanswered, and with the same importance.\footnote{A more readable version can be found at \url{https://docs.google.com/open?id=0B7JbxpuZS0IsRFotZ2VuU0lpcTA}.
Better fonts and more elegant notation in this version.}

That is not what usually happens. Almost always, when we progress in moving to more advanced questions and closer to practical application, we forget that the fundamental concepts and questions remains unanswered or misunderstood. So begins the illusion: we believe that we understand; we believe that the obvious about these concepts is the only relevant truth. And when you think you understand something, there's no point in investigating it any further. For Newton, space and time were self-evident concepts, but just it was not satisfactory to Einstein, and from his investigations emerged the Theory of Relativity.

But that illusion began to fall. Not yet in common sense, but in view of many scholars. The last century was marked by the presence of several lines of thought trying to clarify foundational concepts and questions that had some success, but most of them died before reaching a consensus. We had a crisis in the foundations of mathematics, the linguistic turn in philosophy, questions about the foundations of logic, space and time, and other concepts. But a debate remains stronger than ever before: the interpretation of quantum theory.\parbreak

Quantum mechanics came to shake the paradigms of science. It is a challenge to human understanding. Everyone who can penetrate a little deeper into its mysteries listen to its claim: `So, do you think you truly understand the nature of reality? '. Classical physics is  mathematically and conceptually simpler, and so it does not question our understanding of the fundamental concepts. Our intuitions and world view already provide us with a satisfactory understanding of the meaning of the theory. In this sense, classical mechanics in its present form is at least superficial compared to the quantum theory.

The quantum theory emerged in the usual way as new theories emerge: to explain phenomena which do not admit any explanation according to the previous theory. In that case, the phenomena were related to events taking place at very small scales, of the order of a few atoms, the most well-known have been the emission spectrum of some atoms, the photoelectric effect and the blackbody radiation. In the course of time, the theory took new directions, gaining in scope, sophistication and application, reaching the point of being behind all of modern physics, with the exception of gravitation, and most of the recent technologies. In the book `The Fabric of Reality' \cite{Deutsch98}, Deutch perfectly pictures the relevance of this theory in the actual panorama of science:
\begin{quote}
There are two theories in physics which are considerably deeper than all others. The first is the general theory of relativity, which as I have said is our best theory of space, time and gravity. The second, quantum theory, is even deeper. Between them, these two theories (\ldots) provide the detailed explanatory and formal framework within which all other theories in modern physics are expressed, and they contain overarching physical principles to which all other theories conform. [\ldots] quantum theory, like relativity, provides a revolutionary new mode of explanation of physical reality. The reason why quantum theory is the deeper of the two lies more outside physics than within it, for its ramifications are very wide, extending far beyond physics -and even beyond Science itself as it is normally conceived.
\end{quote}\qparbreak

Its results have already been verified with an unprecedented accuracy, and there is no experiment or unexplained phenomenon that sets doubts on the validity of the theory, nevertheless hundreds of physicists and philosophers all over the world are unsatisfied with the current form of the theory. They do not question the experimental results of the theory, but \emph{the understanding we have about it, the real meaning of the theory, what it tells us about nature and what is the correct interpretation of its mathematical elements}. This is the paradigm shift. There are hundreds of worldwide scientists  whose activities go far beyond attempting to explain phenomena and obtaining new experimental data or theoretical results.

The activity of these researchers goes \emph{beyond the original scope of science}. The question they try to solve goes beyond a scientific problem; it is a true philosophical problem. Von Weizsäcker said this clearly in his book, `The Structure of Physics' \cite[p. 244]{Weiz}: \emph{`The interpretation of physics is a philosophical task. It is a supplementary task, such as we are confronted with it.'}. But in this sense, we have the emergence of a new kind of philosophy or perhaps a return to the original sense of the word, when physics and natural sciences in general, at that time called natural philosophy, were structured on the concepts of philosophy and both were in perfect harmony. As in ancient times, when `science' was done by philosophers, now, this `philosophy' is done by scientists. There is a new philosophy that is not \emph{non-scientific}, but \emph{meta-scientific}. We use the prefix `\emph{meta}' here to denote a concept which is an abstraction of the other, it originates from Greek and means `beyond', `after', `adjacent'. It must be clear that this is not in any way a return to the philosophy and the physics of Aristotle's time, but simply a return to the view of a unity behind philosophy and natural sciences.

I am not saying that the existing Philosophy will be of much help for physicists, and that they should start studying what professional philosophers are doing. I do not believe Philosophy, at this moment, can provide any useful guidelines for science, an opinion that is shared among others, including Weinberg. In his words: \emph{``But we should not expect it [Philosophy] to provide today's scientists with any useful guidance about how to go about their work or about what they are likely to find.''}. I'm not criticizing it in any sense, since I don't think that any professional philosopher of today believes that his work is supposed to provide guiding for physics or even science in general. In fact, physicists have shown to be remarkably independent, along history they have created almost all the necessary tools in order to advance in their fields. They have created new math, new philosophies and even a single building across several countries, the HLC.

I believe any actual definition of philosophy describes only the potential of the socially constructed body of knowledge called Philosophy, not its present situation. For now, I believe philosophy is a word reserved for something greater than we could imagine, but still has not reached its true potential. The situation is not different from Science five hundred years ago. At that time, Science was nothing compared to today's Science, but even so, it deserved respect, not really by what it actually was, but by the effort and seriousness of its members, and the potential it represents.\parbreak

Where I want to get, is the point where physicists realize that the quest for interpreting quantum theory is far deeper and more impacting than we could have ever imagined; with the inherent potential to change not only our understanding of physics, but of philosophy itself. Within this line of thought, Weizsäcker, in the same book  says:
\begin{quote}
 ``Not only is it incompatible with the world view of classical physics, but also with certain positions of classical metaphysics. [\ldots] The present book is based on the conviction that we are dealing with fundamental philosophical progress. According to this conviction, it is not quantum theory that must defend itself before the court of traditional philosophies but those philosophies themselves must stand trial-in itself a philosophical process-with quantum theory in the witness stand.'' \cite[p. 244]{Weiz}
\end{quote}
\vspace{-\baselineskip}\noindent Also Feynman is clearly stating that quantum theory represents nothing less than a new worldview:
\begin{quote}
 ``We have always had a great deal of difficulty understanding the world view that quantum mechanics represents. At least I do, because I'm an old enough man that I haven't got to the point that this stuff is obvious to me. ''\cite[p. 471]{Feynman82}
\end{quote}
\qparbreak

Once again, the one thing many physicists have not realized yet is that only the \emph{pragmatic aspect} of quantum mechanics is really consistent and have reached a level of maturity, in other words, quantum theory is precise \emph{mathematical tool} for calculating the probabilities of the outcomes of an experiment. However, the \emph{true understanding} of the theory and its philosophical conclusions are still open problems. The theory has still not reached its true potential. Feynman's famous phrase: \emph{ ``I think I can safely say that nobody understands quantum mechanics.''} remains valid until the present day! But until when? In this line, Fuchs \cite{Fuchs02} talks about the urgent need to solve this issue, and states that in the thirty years that have passed, from 1972 to 2001, has not had even one year without there being a meeting or conference dedicated to the foundations of quantum theory, and you can go anywhere and you will never find a solution. In that article, he refers to seven different main lines of interpretation. In the encyclopedias\cite {Wikipedia} you can find more than 16 different interpretations, some of them having many different variants. But the situation is somehow hidden because of a declining illusion that there is a `standard interpretation', the Copenhagen interpretation. Many people, because they only had contact with it, think it is the only existing and more consistent one, but even within it there is not a real consensus, since its founders, Bohr and Heisenberg, contradicted each other in key aspects. \emph{Total lack of consensus}, this is how I describe the present situation. \parbreak

In the influential book `The Structure of Scientific Revolutions', the physicist and philosopher Thomas Kuhn argues that \emph{``\ldots proliferation of versions of a theory is a very usual symptom of crisis [in science]''} \cite[p . 71]{Kuhn}. By considering each new interpretation as a different version, we can see that there is a subtle crisis, or perhaps, just hidden under the title of 'interpretation of quantum theory', which is beginning in our historical moment.

After all this discussions, now we can get to its climax. I am about to promote the quest of interpreting quantum theory to a \emph{whole new level}, that will give us a new insight on how or what the interpretation of quantum theory is supposed to be. The main point is that most, if not all, of its paradoxes are not paradoxes in the sense of logic, that is, contradictions with the theory itself;  but only a disagreement with what is expected to be. The paradoxes are simply aspects of the interpretations that at least appears to be incompatible with our intuitions or our worldview. In face of it, the question we should be asking is not simply `how to interpret quantum theory', but
\begin{quote}
What is the worldview that makes quantum theory understandable?
\end{quote}
\vspace{15mm}

\section{Formulation of Physical Theories}

In the formulation of a physical theory, the discussions, conclusions and most of the interpretation are written in natural language. The interpretation is not something outside, separated of the theory, but clearly outside its mathematical formalism, and so is something that makes extensive use of natural language. So, when we say `an interpretation of a theory' we actually mean `an interpretation of the mathematical formalism of the theory', since the interpretation itself is part of the theory. In fact, the existence of a physical interpretation is what distinguishes theoretical physics from pure math.

Without a minimal interpretation nothing can be said or, maybe, even understood about the theory. It becomes just an abstract mathematical theory, a list of equations without any other meaning except what math says by itself. If we can't say anything about the theory, we can't relate to anything else neither discuss, conclude anything or even find an application to it. That is just like that. Both math and natural languages are necessary and a great deal of our ability to understand a physical theory is related to our means of talking about the theory. One of the explanations for it is that most of our thinking requires the use of language, therefore, if you can not speak, you probably can't think either.\parbreak

Once there is not a singular, consensual interpretation of quantum theory, there is not an unique quantum theory. What we have is, being optimistic, just a unique mathematical formalism. Trying to remove the interpretation from the theory is like reducing a physical theory to a purely mathematical theory. Quantum theory at its present form is essentially incomplete. \parbreak

But in any physical theory, is in its non-mathematical part where you get the least consensus and most of the ambiguities. If you look though history, you will see that the equations within the textbooks of physics have changed very little compared to the text within the book. In fact, you may find that you disagree with most of what Newton and other physicists said originally about their works.

Most of the precision and power of contemporary physics come from the extensive use of math, and, if you remove all the math from it, you may notice that you are back to the physics we had five hundred years ago. What I mean is that, if you take any graduated textbook of physics and remove all its mathematical equations and then try to read what was left; you will see that, although you may understand the idea and the fundamental concepts you cant find almost no application or quantitative result; actually, physics becomes a purely qualitative discipline. What could you do with classical mechanics without using that $\vec{F}= m \vec{a}$ ? Well, we could still philosophize about it\ldots \parbreak

\section{Interpretation of a Physical Theory}

Could we  answer what is an interpretation of a physical theory now? The most important is that it is something which allow us to talk about the theory using natural language so that we could draw conclusions and relate it to other things. Therefore I'll give the following definition:
\begin{quote}
The interpretation of a mathematical formalism is what establishes the connection between the mathematical elements or equations and the natural language used to talk about the theory.
\end{quote}

Once we have an interpretation, we are able to ``read'' the equations or other mathematical elements and then we can say, \emph{in plain words}, what the theory says that is valid. We can also define what it means to interpret a mathematical statement:
\begin{quote}
To interpret a mathematical statement is to say in natural language the meaning of it.
\end{quote}

Thus, the interpretation of the mathematical formalism allow us to interpret each mathematical statement of the theory, and that means to reveal the meaning of each mathematical statement. We could say that the interpretation of a mathematical statement is an objective statement of the theory. Given an interpretation, all the objective statements follow directly from the mathematical formalism.

The use of the word `objective' here is to emphasize that it is not some kind of philosophical statement that could change in the future. The objective statements are almost as reliable as a mathematical statement; however this notion has not been used yet and we are not precisely able to distinguish in our current theories what is an objective statement and what is not. \parbreak

Until this point everything is clear and precise, but many of us can't stop just here\ldots

\section{The Obscurity of our Fundamental Concepts}

After this point we have no more the support and clear guidance from math. Farther you get from it, more you enter into a no one's land, a place of obscurity where anyone may give a different meaning to your words and many times, neither you know the true meaning of the words you are using. Any communication becomes limited since the consensus exists only in the superficial level of the words, not in their full depth. The meaning of these words is just like the weather: when you look to the whole it looks homogeneous, but, closer you look, more you see how it differs from place to place. This is where philosophers feel at home, and from where scientists try to stay as far as they can.\parbreak

For now, this is where all of our fundamental concepts live. There is not a clear, consensual meaning that reveals the full depth of any of them, and our words easily betray us. \emph{World, universe, reality, to be real, existence, to exist, object, state, proposition, property, quantity, value, truth, observer, knowledge, information, probability, determinism, time, space,} \ldots\  are just a few examples of them. \emph{Any attempt} to define these words using \emph{natural language} is fruitless. One word can only be defined in terms of the others, and so to define a fundamental word you need other fundamental words and sometime this will become circular. The most common solution is to define such words using another not so deep word that people believe that they understand. However, this is not a definition. We may use bricks to build a house, not the opposite. The same holds for giving meaning to words.

The (partial) meaning each one has about these concepts forms the basis of each one's worldview. The existing consensus on their meaning forms our actual worldview. There is not a single worldview.\parbreak

But when it comes to quantum theory, unlike classical mechanics, it is so deep that it is almost impossible to explain it without entering in this domain of obscurity where our fundamental concepts actually are. And this is the greatest challenge of interpreting quantum theory: \emph{there can be no really consistent connection between such a beautiful mathematical formalism and such vague concepts lacking any clear meaning.} There is no way the proper interpretation of quantum theory could fit in the plurality and obscurity of our current worldviews. \emph{Quantum theory requires a new worldview.} \parbreak

Once more we got into a climax: 
\begin{enumerate}
\item Quantum theory is part of a \emph{final theory.}
\item A final theory is not an approximation to a better theory.
\item A final theory is not based on an unknown deeper theory.
\item A final theory is free of any philosophical preconception.
\item A final theory \emph{defines} a set of fundamental concepts.
\item The interpretation of a final theory gives mathematical precision and significance to a small subset of our natural languages.
\item The final theories define the main foundations of a unified consistent worldview. They are \emph{the only possibility of it.}
\end{enumerate}

\chapter{Final Theories}

We will propose a theory of this kind. This section is to give us some guidelines such that we could recognize a final theory and be able to understand its meaning and maybe even improve it. The theory contains a new formulation of the mathematical apparatus of quantum theory that is entirely axiomatic, generalizes many notions and add lots of useful mathematical tools. It is remarkably simple and elegant as it is supposed to be, and might be very insightful even if you do not agree with some its conclusion and interpretations.

From the mathematical point of view, it presents an unified view on many different mathematical notions and structures; and might be the richest mathematical theory of all times. This formulation tries to unveil the most of the beauty hidden in the math of quantum theory, however, further advances in this direction may require advances of our understanding and possibly on the power of our current math. 

The interpretation of this theory takes an entirely different approach.
Almost all of its interpretation is a direct interpretation of each of its mathematical statements, so it is primarily composed of objective statements. The theory is a theory of knowledge and truth, so it has both a subjective or epistemic aspect (knowledge), and an objective or ontic aspect (truth), all them in a complete harmony. \parbreak

It is not the first time someone talks about final theories. Weinberg states that
\begin{quote}
``This theoretical failure to find a plausible alternative to quantum mechanics... suggests to me that quantum mechanics is the way it is because any small change in quantum mechanics would lead to logical absurdities. If this is true, quantum mechanics may be a permanent part of physics. Indeed, quantum mechanics may survive not merely as an approximation to a deeper truth, in the way that Newton's theory of gravitation survives as an approximation to Einstein's general theory of relativity, but as a precisely valid feature of the final theory.'' \cite{Weinberg94}
\end{quote}

In a similar fashion, also Von Weizsäcker says that:
\begin{quote}
``It [quantum theory] is a closed theory in the sense of Heisenberg. It cannot, so it would seem, be further improved via small modifications. But can one imagine the progress of science passing through an infinite succession of closed theories? If not, then one of them must be the last, the final one. Why not quantum theory? \ldots '' \cite[p. 311]{Weiz}
\end{quote}

\section{Closed Theories}

In common to both views, there is the notion of a closed theory that dates back to Heisenberg: ``By a closed theory we mean a system of axioms, definitions and laws, whereby a large field of phenomena can be described, that is mathematically represented, in a correct and non-contradictory fashion''\cite[p.123]{Heisenberg72}. But this definition is still not complete.

For him \cite{Heisenberg58}, a closed theory is also a self-contained system of concepts that can be represented by mathematical symbols with equations representing the different relations between different concepts. So, for a theory to be considered closed, it is necessary not only for its mathematical formalism to satisfy certain properties but also its conceptual framework, and hence its interpretation.

According to Heisenberg, classical mechanics, thermodynamics, electromagnetism and quantum mechanics are closed theories. However, none of them is actually being formulated in such a formal axiomatic basis as he describes, although it is not hard to believe that they \emph{could} be axiomatically formulated in a way just as formal as our most rigorous mathematical theories.

To make theses notions clearer, lets state what means for a theory to be in a closed form or be in a closed formulation. The form of the theory is the way it is actually being formulated, and a theory is a closed theory if it could be formulated in a closed form.

A theory is in a closed form when its formulation satisfies:
\begin{enumerate}
\item The mathematical formalism of the theory is a formal mathematical theory; in other words, it is composed by an axiomatic system and all its derived theorems. The axiomatic system is a set of axioms that defines the mathematical formalism.

\item The interpretation of the theory provides means of reading in plain language each mathematical statement relevant to the theory. The meaning of each concept used in the interpretation is defined within the theory.

\item Each statement about the theory is the interpretation of some of its theorems or axioms. Each statement is an objective statement. There is no room for philosophical preconceptions.

\end{enumerate}

The only way of changing a closed theory is changing its interpretation or some of its axioms. However, in any axiomatic system, the change of a single axiom is a dramatic change and the resulting theory will be an entirely different theory, and in most cases a fruitless one. If you change the interpretation, the new set of concepts will be isomorphic to the old set of concepts, and so there is no real change.  Therefore, any change in a closed theory leads you to a totally different theory, and you can't say that you just improving it or making a small modification. That's why we say that a closed theory is not open to modifications. \parbreak

Since a closed theory defines a consistent set of concepts, that is, it gives meaning to a set of words that represents concepts, we could state that it defines part of a worldview. The worldview defined by a closed theory is precisely the one that makes the theory understandable and avoids any paradoxes. \parbreak

Now it is not difficult to realize that quantum mechanics is not yet in its closed form, even we believing that it actually is a closed theory. We can also see how much we could benefit if we could provide a closed formulation for it.

\section{Final Theories}

Within the domain of closed theories, one may find a theory that is so general, that has such a limited number of axioms that the theory is not making a single assumption about the world. It is the most general and fundamental theory possible within its field. It cannot be succeeded by another more general theory since this would imply that there is a theory that makes fewer assumptions. We call such theory a \emph{Final Theory.}\parbreak

We might have final theories for each branch of knowledge, that is, one for physical sciences, biology, economy and even psychology and social sciences, given that each of those branches requires a new dimension of concepts. The final theory that unifies the final theories of each of those branches of knowledge deserves to be called the true \emph{Theory of Everything.}\parbreak

Within a branch of knowledge there is no other theory that is deeper or more fundamental than its final theory. Since it defines the concepts, it cannot be based on another unknown theory. The final theory provides the conceptual and mathematical formalism over which all the other theories can, and should ideally be built over. The final theory is what is in common to all the closed form theories within that branch of knowledge, and so, only closed theories can be explicitly formulated over the final theory.\parbreak

The validity  of a final theory might be judged by its aesthetics, but not by common sense or philosophy. Actually is the opposite that should ideally occur. We should change (in fact correct) our worldview in order for it to be compatible with the theory, since the contrary simply cannot happen. We cannot change its mathematical formalism. However, there are other ways of interpreting it. Actually this give us a criterion for a good interpretation: the ideal interpretation is the one that is the most compatible with our worldview, in other words, it is the one that requires minor changes to it.

Now we can state that the correct interpretation of the mathematical formalism of a final theory (or any closed theory) is the one that requires the minimum number of changes to or current worldview. And also, between a list of closed theories with the same mathematical formalism, the best one is the one that is the most compatible with our current worldview.

But what it means for a theory to be compatible with our worldview? A final theory or any closed form theory gives meaning to a set of words, concepts or expressions of natural language. Our worldview also gives meaning to expressions, concepts and words, but we cannot compare the meanings itself. The only way, is to compare how two or more concepts or expressions are related according to our worldview, and according to the theory. That is, we compare the statements in ordinary language about an expression or concept according to our worldview, and according to the theory. But the problem is that final theories are so deep that they deal with concepts with little or no consensus neither in common sense neither in philosophy.\parbreak

Also, a final theory cannot be judged by empirical experiments. It is not a falsifiable theory. In fact it may determine what it means for a theory to be falsifiable, which is for now a purely philosophical concept. So, in a sense, a final theory is not a scientific theory, but the fundamental framework over which scientific theories can, and ideally should, be built over.\parbreak

Given the depth of a final theory, one may notice that it contains two extremes.  Its mathematical part looks like pure and abstract math. And its statements in ordinary language look like purely philosophical ones. Together they form something new, something that belongs, not to a specific subject, but to the unity of them. The existence of a final theory of certain branch of knowledge is the ultimate proof, or even the manifestation itself, of the underlying unity of all its subjects. \parbreak

A final theory is the ultimate expression of the ideals of beauty, simplicity and economy. Each theory that is explicitly integrated into the final theory, has reached its full potential in term of beauty, simplicity and clarity. And so, the existence of final theories explains why aesthetics have been such an important guiding for physicists. When we are navigating over the dark and unexplored areas of the unknown, and we still don't know the meaning of our equations, the ideal of beauty is our only guide.

Once we find a final theory, it will give us no knowledge about the world, but a fundamental understanding of it, and a framework where our knowledge can be organized. No matter how God created the world, no matter what have we done with it, the final theories still hold. They are immutable, unique and a manifestation of the unity. They represent the possibility of synthesis of all our knowledge and understanding. \parbreak

Each final theory represents a fragment of the mind of God.

\section{The Final or Fundamental Theory of Physics}

Of all possible final theories, the Final Theory of Physics (FTP) is the first and the most fundamental one, and hence it can be called the Fundamental Theory of Physics. Given the level of complexity of all the other branches of knowledge and the fact that all our closed theories belongs to physics, there is no way we could discover other final theories before discovering the one of physics.

Given the role played by physics in the panorama of all the physical sciences, the fundamental theory of physics will be also the final theory of physical sciences. This theory represents the possibility of unification of all the subjects within physics or even physical sciences as whole. The FTP is itself the expression of the Unity of Physics!\parbreak

The FTP will not contain any of our existing theories, but it is the basis for them. So it might contain part of our current theories, but not entirely any of them. In fact, it contains the basic elements and the formalism of our quantum theory.\parbreak

In the next chapter the foundations of FTP will be formulated. We mean `the foundations' just because the theory might be extended while being the FTP, and because its mathematical formulation can be improved.

The FTP answers the question ``what is the nature of reality?'', however, the answer can only be expressed in the language of mathematics. The theory finally defines some fundamental concepts and notions, like reality, state, truth, falsity, knowledge, probability, possibility, necessity, among others.\parbreak

The theory contains only a fraction of the present time quantum mechanics, and that is about the state of the system, Hilbert space and probability. The time evolution, commutation relations, Hamiltonian and other notions are not mentioned and probably include other assumptions that declassify them as part of the final theory.

Although it uses the mathematics of quantum theory, the FTP does not invalidates classical mechanics, in fact, classical mechanics can be integrated to it, as the other closed theories.\parbreak

The foundations of all the mathematical formalism of quantum theory are formulated in a new unified and completely axiomatic approach that is essentially different from any other formulation of a physical theory up to this time.

We have added extensions to the usual language of mathematics providing new very powerful tools. Many useful mathematical notations, definitions and relations are added to the existing formalism of quantum theory in an attempt to reveal its true beauty.\parbreak

The interpretation of the theory addresses most of the controversial aspects of today's quantum theory. The interpretation takes the form that was mentioned here, and so, it is different in nature from every previous interpretation, and deeper, much deeper. However, no extensive comparison of the proposed interpretation and the previous ones is actually done.

The theory gives the meaning of the word `reality', and once we understand it, many old questions become simply meaningless, that is, to ask them means that you do not understand the meaning of the questions itself. The interpretation of this theory provides the basic tool for dissolving considerable philosophical problems of quantum theory and even of philosophy itself.\parbreak

This work inaugurates a new way of formulating physical (closed) theories. All the formulation is subdivided in a unique structure: we have few sections of ordinary text, and the rest is composed by specific sections of the types: \textsc{Formalism}, \textsc{Definition}, \textsc{Interpretation} and \textsc{Statement}. Each section represents a different aspect of the theory:

\begin{enumerate}
\item In a \textsc{Formalism} section we define special elements of the mathematical apparatus, like functions and binary relations.

\item In a \textsc{Definition} section we have an explicit definition of a special set of the theory. A special set means a set that defines a concept.

\item In an \textsc{Interpretation} section we explain how should be the connection between ordinary language and a specific part of the formalism; that is, we give names to mathematical objects or determine how you should read a mathematical expression.

\item In a \textsc{Statement} section we present a theorem, that is, some result that follows from the definitions or an axiom. In each Statement we have a sentence written in ordinary language and the corresponding mathematical equation. The sentence is an objective interpretation of the following equation. \parbreak
\end{enumerate}

No previous knowledge of quantum mechanics is required for the understanding of the theory; and it is essentially self-contained. It can be understood by non-physicists, but the reader should be versed in the language of mathematics. If you can't understand the math, all that you will read will sound like essentially philosophical, something like a philosophical treatise since no argumentation is provided (in ordinary language).

If you are a physicist, by the end you will be asking when the physics we know really begins. Well, it begins when we make the physical assumptions that arise each subject. To know precisely what are those assumptions will be one of the greatest contributions to physics. Those assumptions will be exactly the ultimate principles of each subject of physics, \emph{the fundamental principles of physics.}


\chapter{The Mathematical Formalism}

\section{Mathematical Preliminaries}
 The characterization of the mathematical formalism is basically the definition o a few operations, relations and sets, and the description of some of its most important properties.

\subsection{Extending the Language of Mathematics}

Since we gonna use lots of definitions of sets, it will be interesting to define new useful notations. We will use ``iff'' as an abbreviation for ``if, and only if,''.

Consider the sets \set{A}, \set{B}, \set{C}, \ldots, and a function $f$, then we define:
\[
f(\set{A}, \set{B}, \set{C},\ldots) \equiv \sdef{f(a,b,c,\ldots) : a \in \set{A}, b \in \set{B}, c \in \set{C}, \ldots},
\]
that is, a function applied to sets is the set of all the possible results of application of the function on elements of the sets. Therefore, we have:
\begin{align}
\Aset + \Aset &\equiv \sdef{a+b : a \in \Aset, b \in \Aset}\\
\Aset  \Aset &\equiv \sdef{a b : a \in \Aset, b \in \Aset}
\end{align}

An important property is that
\[
(\forall a \in \Aset, f(a) \in \Bset) \then f(\Aset) \subseteq \Bset,
\]
With it we can easily define what it means to a set \Aset\ be closed under a certain operation $f$. Depending on the arity of $f$ it is:
\begin{align}
f(\Aset) &\subseteq \Aset\\
f(\Aset,\Aset) &\subseteq \Aset\\
\vdots &\notag
\end{align}

In a similar way we define an analogous notation for relations. Let \Aset\ and \Bset\ be sets, and $R$ a binary relation defined on this sets, then we define
\[
\Aset \mathrel{R} \Bset \equiv \sdef{(a,b) : a \mathrel{R} b, a \in \Aset, b\in \Bset}
\]
Two important property are that:
\begin{align*}
\forall \Aset, \Bset, R: \qquad \Aset \mathrel{R} \Bset &\subseteq \Aset \times \Bset,\\
\forall a \in \Aset, b \in \Bset: \quad a \mathrel{R} b &\iff (a,b) \in \Aset \mathrel{R} \Bset
\end{align*}

When useful, we may also use the prefix notation instead of the usual infix notation, therefore, for any binary operation $+$,
\[
+(a,b) \equiv a + b,
\]
and, for example: 
\[
+(\Aset \mathrel{R} \Aset) \equiv \sdef{a+b : a \mathrel{R} b,\, a \in \Aset, b\in \Bset}
\]
read as the sum of every two $R$-relates members of \Aset.

\subsection{Concepts}

We will be frequently dealing with sets that are somehow special, like the set \Rset\ of the Real Numbers. In fact the set \Rset\ defines the concept of a real number, so for example, \Xset\ is a set of real numbers iff $\Xset \subseteq \Rset$. And to be a real number means to belong to \Rset, that is, $x$ is a real number iff $x \in \Rset$. With these ideas we can connect the notion of concept with the notion of set:

\textit{Let $X$ be a concept name, then, to say that the set \Xset\ is concept of $X$, means that, an element $a$ is a/an $X$ iff $a \in \Xset$; or that, $a$ is a/an $X$ iff $a$ belongs to the concept of $X$.}

This is important to remove ambiguities in ordinary language, for example, what is the set of real numbers? Any set $\Aset \subset \Rset$ is a set of real numbers, in some contexts the article `the' may lead us to think of \Rset, but in others it may refer to a previously mentioned set and so there is a risk of ambiguity. To clarify it,

\textit{Let $X$ be a concept name and \Xset\ the concept of $X$, then to say that \Aset\ is a set of $X$s means that $\Aset \subseteq \Xset$.}

Concepts are sets with an important meaning within the theory. Note that we wont be talking about words or concept names, but concepts itself. So the theory defines the concepts and the interpretation associate the concepts with the associated name. So a translation of this work may change the names of the concepts, not the concepts itself. 

Within our framework there is an universal set \ets\ and every meaningful element lies in \ets. We call \ets\ the Universal Category.

The Universal Category defines the concept of element. Every member of the universal category is called an element.

There is no element that is also a set. Every concept we will define is a subset of the universal category.

\subsection{Categories}

 There are also special concepts that we call categories. Categories are the most general concepts that couldn't be defined otherwise. They can be recognized by which other elements or concepts appears in their definition.
\begin{quote}
 A concept in which in its definition appears only the distinct elements of the universal category is called a category.

A concept in  which in its definition appears only the distinct elements or other categories is called a category.
\end{quote}

We wont consider anything else being a category. This notion of category is important because categories are kinds of concepts that do not depend on anyway on how the `world out there' is, they cant be otherwise. As we will see, the concept of proposition is a category, but the concept of truth is not. As expected, the truth depends on how the world is.

We are removing any need for a philosophical baggage, thus, everything can be defined mathematically.


\pagebreak
\section{The Universal Category \ets}

The universal category \ets\ posses the algebraic structure of a star ring with identity \tuple{\ets,[+,\cdot, \st],\sdef{0,1}}, that is, \ets\ is closed by two binary operations called addition ($+$) and multiplication ($\cdot$), and an unary operation ($\st$), contains two distinct members, zero ($0$) and one ($1$), and satisfies:

\begin{equation}
\sdef{0,1} \subset \ets\dsp \ets\s = \ets\dsp
\ets+\ets = \ets\dsp \ets\cdot\ets = \ets,
\end{equation}
$\forall a, b, c \in \ets:$\vspace{-\baselineskip}
\twocolumneq{
\begin{align}
a+b &= b+ a \\
a+(b+c) &= (a+b) + c \\
a+0 &= 0 \\
\exists (-a) \in \ets:\; a + (-a) &= 0\eqspace
a\cdot(b\cdot c) &= (a\cdot b) \cdot c \\
a\cdot 1 = 1\cdot a &= a 
\end{align}
}
{
\begin{align}
a\cdot(b + c) &= a\cdot b + a\cdot c \\
(a+b)\cdot c &= a\cdot c +b\cdot c \eqspace
(a+b)\s &= a\s + b\s \\
(a\cdot b)\s &= b\s\cdot a\s \\
a\s\s &= a \\
\notag
\end{align}
}

The subtraction operation ($-$) is defined using the additive inverse by
\begin{cequation}{ a, b \in \ets}
a - b \eqd a + (-b).
\end{cequation}

From now on, will omit the multiplication symbol ($\cdot$).

\begin{interpretation}
The [sum of $a$ and $b$] or, [$a$ plus $b$], is denoted by $a + b$. 
\end{interpretation}
\begin{interpretation}
The [product of $a$ and $b$] or, [$a$ times $b$], is denoted by $a b$. 
\end{interpretation}
\begin{interpretation}
The [adjoint of $a$] is  denoted by $a\s$. 
\end{interpretation}

\begin{statement}
Two elements are equal if, and only if, their adjoint are equal.
\begin{cequation}{ a,b \in \ets}
a = b \iff a\s=b\s
\end{cequation}
\end{statement}

\subsection{Compatibility Relation}
\begin{formalism}
We define the compatibility or commutativity relation ($\cpt$) between any two elements by
\begin{cequation}{ a,b \in \ets}
a \cpt b \rld a b=b a,
\end{cequation}
and between any number of elements by:
\begin{cequation}{ a,b,c \in \ets}
a \cpt b \cpt c \rld (a \cpt b\tand a\cpt c\tand b\cpt c).
\end{cequation}
\end{formalism}

\begin{interpretation}
[$a$ is compatible with $b$], or [$a$ and $b$ are compatible] iff $a \cpt b$ holds. 
\end{interpretation}

\begin{interpretation}
The elements [$a$, $b$ and $c$ are compatible] iff $a \cpt b \cpt c$ holds.
\end{interpretation}

\begin{statement}
Every element is compatible with itself. The compatibility relation is reflexive.
\begin{cequation}{ a \in \ets}
a \cpt a. 
\end{cequation}
\end{statement}

\begin{statement}
If $a$ is compatible with $b$ then $b$ is compatible with $a$. The compatibility relation is symmetric.
\begin{cequation}{ a, b \in \ets}
a \cpt b \iff b \cpt a.
\end{cequation}
\end{statement}

\begin{statement}
If two elements are compatible then their adjoint are also compatible.
\begin{cequation}{ a,b \in \ets}
a \cpt b \iff a\s\cpt b\s.
\end{cequation}
\end{statement}

\subsection{Self-adjoint Elements}

\begin{interpretation}
We say that an element [$a$ is self-adjoint] iff $a=a\s$ holds.
\end{interpretation}

\begin{statement}
The product of two self adjoint elements is self-adjoint iff they are compatible
\begin{cequation}{ a,b \in \ets}
a=a\s\s \tand b=b \then \big (ab=(ab)\s \iff a \cpt b\big )
\end{cequation}
\end{statement}

\subsection{The Null-Product Relation}

\begin{formalism}
We define another binary relation called null-product relation $\ctrd$ by
\begin{cequation}{ a, b \in \ets}
a \ctrd b \rld( ab=0).
\end{cequation}
\end{formalism}

\begin{statement}
Between two self-adjoint elements the null-product relation is symmetric.
\begin{cequation}{ a, b \in \ets}
a\s=a \tand b\s=b \then \big(a \ctrd b \iff b \ctrd a\big).
\end{cequation}
\end{statement}

\subsection{Idempotent Elements}

\begin{interpretation}
We say that an element [$a$ is idempotent] iff $a^2=a$ holds.
\end{interpretation}

\begin{statement}
The product of two compatible idempotent elements is also idempotent:
\begin{cequation}{ a,b \in \ets}
\big(a^2 = a\tand  b^2=b \tand a \cpt b\big)\;\; \then \;\;(ab)^2 = ab.
\end{cequation}
\end{statement}

\begin{statement}
If an element is idempotent then its adjoint is also idempotent.
\begin{cequation}{ a \in \ets}
a^2=a \iff (a\s)^2=a\s.
\end{cequation}
\end{statement}

\subsection{Invariance Relation}

\begin{formalism}
We define the invariance binary relation (\pleq) between any two elements by:
\begin{cequation}{ a, B \in \ets}
a \pleq B \rld B a B\s = a.
\end{cequation}
\end{formalism}

\begin{interpretation}
An element [$a$ is invariant under $B$] iff $a \pleq B$ holds.
\end{interpretation}

\begin{formalism}
We also define the opposite of this relation:
\begin{cequation}{ a, B \in \ets}
B \pgeq a \rld a \pleq B.
\end{cequation}
\end{formalism}

\begin{interpretation}
We say that [$U$ lets $a$ invariant] iff $U \pgeq a$ holds.
\end{interpretation}

\begin{statement}
Any element is invariant under $1$.
\begin{cequation}{ a \in \ets}
a \pleq 1.
\end{cequation}
\end{statement}

\begin{statement}
Every element lets $0$ invariant.
\begin{cequation}{ a \in \ets}
a \pgeq 0.
\end{cequation}
\end{statement}

\begin{statement}
If element $a$ is invariant under another element $B$ then its adjoint $a\s$ is also invariant under $B$.
\begin{cequation}{ a, B \in \ets}
a \pleq B \iff a\s \pleq B.
\end{cequation}
\end{statement}

\section{Numbers}

\subsection{The Null Category or Null Set \ctrs}

The distinguished element zero defines the null category, or null set, \ctrs:
\begin{edefinition}
\ctrs \eqd \sdef{0},
\end{edefinition}
which forms the trivial star ring:
\begin{equation}
\ctrs + \ctrs = \ctrs \dsp \ctrs \ctrs = \ctrs \dsp -\ctrs = \ctrs\dsp \ctrs\s=\ctrs
\end{equation}

\begin{formalism}
For any set $\Xset \subseteq \ets$, we define $\Xset_*$ to be:
\begin{cequation}{\Xset \subseteq \ets}
\Xset_* = \Xset\backslash\ctrs.
\end{cequation}
\end{formalism}

\subsection{The Category of Booleans \bools}

The distinguished elements $0$ and $1$ defines the boolean category or category of booleans \bools:
\begin{edefinition}
\bools \eqd \sdef{0,1},
\end{edefinition}
which is closed under adjoint operation and multiplication:
\begin{equation}
\bools\s = \bools\dsp \bools \bools = \bools.
\end{equation}

There is an important property that the boolean elements are compatible with every element of \ets:
\begin{cequation}{a \in \ets\bsp \alpha \in \bools}
\alpha \cpt a.
\end{cequation}

\subsection{The Category of Natural Numbers \Nset}

We define the category of natural numbers \Nset\ to be such that \Nset\ is the smallest set containing the boolean category which is closed under addition and multiplication:
\begin{edefinition}
\Nset \eqd \sum \bools.
\end{edefinition}

The summation without index is used to express that any finite sum of boolean elements is in \Nset. That is equivalent to a recursive definition of the natural numbers and satisfies:
\begin{equation}
\begin{array}{c}
\bools \subset \Nset  \subset \ets,\\
 \Nset + \Nset = \Nset\dsp \Nset \Nset = \Nset\dsp\Nset\s = \Nset,\vspace{5pt}\\
\end{array}
\end{equation}%
\nobreak \vspace{-20pt}\nobreak
\begin{cequation}{ \alpha \in \Nset\bsp a \in \ets}
\begin{aligned}
\alpha &\cpt a\\
\alpha\s &= \alpha. 
\end{aligned}
\end{cequation}

The natural numbers are ordered by a total order relation $\leq$ (less than or equal to) and equivalently by a strict order $<$ (less than), with $a < b \iff a \leq b \tand a \neq b$:
\[
0 < 1 < 2 < 3 \cdots
\]
that can be defined by:
\begin{cequation}{a,b \in \Nset}
a\leq b \iff \exists n \in \Nset: a+n = b.
\end{cequation}

\subsection{The Category of Integers \Zset}

The category of integers \Zset\ is defined as the minimal extension of the natural numbers in which every element posses its additive inverse, or that is closed under subtraction:
\begin{edefinition}
\Zset \eqd \Nset \un (-\Nset),
\end{edefinition}
and we have that:
\begin{equation}
\begin{array}{c}
\bools \subset \Nset \subset \Zset \subset  \ets,\\
 \Zset + \Zset = \Zset\dsp \Zset \Zset = \Zset\dsp\Zset\s = \Zset\dsp -\Zset = \Zset,
\end{array}
\end{equation}
\nobreak \vspace{-20pt}\nobreak
\begin{cequation}{ \alpha \in \Zset\bsp a \in \ets}
\begin{aligned}
\alpha &\cpt a\\
\alpha\s &= \alpha. 
\end{aligned}
\end{cequation}

\subsection{The Category of Rationals \Qset}

We define a multiplicative inverse $\alpha^{-1}$ of $\alpha$ to be such that 
\[
\alpha \alpha^{-1} = \alpha^{-1} \alpha = 1,
\]
and so we extend the integers in order to be closed under the division operation defined by:
\[
\frac{a}{b} \eqd a b^{-1}
\]
The category of rational numbers is then defined by:
\begin{edefinition}
\Qset \eqd \frac{\Zset}{\Zset_*}.
\end{edefinition}

The rational numbers satisfy:
\begin{equation}
\begin{array}{c}
\bools \subset \Nset \subset \Zset \subset  \ets,\\
 \Qset + \Qset = \Qset\dsp \Qset \Qset = \Qset\dsp\Qset\s = \Qset\dsp -\Qset = \Qset\dsp (\Qset_*)^{-1}=\Qset_*, 
\end{array}
\end{equation}%
\nobreak \vspace{-20pt}\nobreak
\begin{cequation}{ \alpha \in \Qset\bsp a \in \ets}
\begin{aligned}
\alpha &\cpt a\\
\alpha\s &= \alpha. 
\end{aligned}
\end{cequation}

This category posses an algebraic structure known as an ordered field.

\subsection{The Category of Real Numbers \Rset}

The real numbers extends the rational numbers maintaining the same algebraic structure, that is, they are also an ordered field. However they cannot be  defined explicitly like the previous categories of numbers, and we have to take an axiomatic approach.

The category of real number can be axiomatically defined by the properties:

\begin{enumerate}
\item \Rset\ is a field composed of self-adjoint elements compatible with every element of \ets:
\begin{gather}
\begin{array}{c}
\bools \subset \Rset \subset \ets,\\
 \Rset + \Rset = \Rset\dsp \Rset \Rset = \Rset\dsp -\Rset = \Rset \dsp (\Rset_*)^{-1}=\Rset_*, \\[5pt]
\begin{aligned}
\forall \alpha \in \Rset\bsp a \in \ets:\quad \alpha &\cpt a \text{\hphantom{$\forall \alpha \in \Rset, \forall a \in \ets\dsp $}}\\
\alpha\s &= \alpha. 
\end{aligned}
\end{array}
\end{gather}

\item \Rset\ is ordered:

It posses a total order relation $\geq$ satisfying:
\begin{subequations}
\begin{gather}
a \geq b \tand b\geq a \then a = b\\
a \geq b \tand b \geq c \then a\geq c\\
a \geq b \tor b \geq a
\end{gather}
which is compatible with the operations of addition and multiplication:
\begin{gather}
a \geq b \then a+ c \geq b+c\\
a \geq 0 \tand b\geq 0 \then a b \geq 0
\end{gather}
\end{subequations}
\item Every non-empty subset of \Rset\ with an upper bound in \Rset\ has a least upper bound in \Rset.

\end{enumerate}
\vspace{10pt}

The real numbers contains all the previous categories of numbers:
\begin{equation}
\bools \subset \Nset \subset \Zset \subset \Qset \subset \Rset,
\end{equation}
and satisfies the property:
\begin{cequation}{ \alpha \in \Rset}
\alpha^2 \geq 0\csp \alpha^2 = 0 \iff \alpha = 0.
\end{cequation}

\subsection{The Category of Complex Numbers \Cset}

The complex numbers are defined to extend the real numbers by possessing the same closures, that is, they are a also a field, however thy do not posses any ordering relation, but satisfies:
\begin{cequation}{\alpha \in \Cset}
\alpha \alpha\s \geq 0.
\end{cequation}

The category of complex numbers is then defined as
\begin{edefinition}
\Cset \eqd \Rset + i \Rset,
\end{edefinition}
in which $i$ is the imaginary unity defined by:
\begin{cequation}{ a \in \ets}
a \cpt i\csp i^2 = -1\csp i\s = -i.
\end{cequation}

The category of complex numbers satisfies:
\begin{equation}
\begin{array}{c}
\bools \subset \Nset \subset \Zset \subset \Qset \subset \Rset \subset \Cset \subset \ets,\\
 \Cset + \Cset = \Cset\dsp \Cset \Cset = \Cset\dsp -\Cset = \Cset\dsp(\Cset_*)^{-1}=\Cset_*,
\end{array}
\end{equation}
\nobreak \vspace{-20pt}\nobreak
\begin{cequation}{ \alpha \in \Cset\bsp a \in \ets}
\begin{aligned}
\alpha &\cpt a.
\end{aligned}
\end{cequation}

\section{Propositions}

\subsection{The Boolean Operations}

\begin{formalism}
We define the main boolean operations, conjunction (\pand), disjunction (\por) and negation (\pnot) by:
\begin{align}
\forall a, b \in \ets: \quad a \pand b &\eqd ab,\\
a \por b &\eqd a + b -ab,\\
\pnot a &\eqd 1- a.
\end{align}
\end{formalism}

\begin{interpretation}
The [conjunction of \prp{a}\ and \prp{b}] is denoted by $\prp{a}\pand \prp{b}$.
\end{interpretation}
\begin{interpretation}
The [disjunction of \prp{a}\ and \prp{b}] is denoted by $\prp{a}\por \prp{b}$.
\end{interpretation}
\begin{interpretation}
The [negation of \prp{a}] is denoted by $\pnot \prp{a}$.
\end{interpretation}

\subsection{On the Boolean Category}

The boolean category is composed of compatible self-adjoint idempotent elements:
\begin{cequation}{ \alpha,\beta \in \Bset}
\alpha\s = \alpha\csp \alpha^2 = \alpha\csp \alpha \cpt \beta.
\end{cequation}

\begin{statement}
The conjunction of two boolean elements is a boolean element. The disjunction of two boolean elements is a boolean element. The negation of a boolean element is a boolean element.
\begin{equation}
\bools \pand \bools = \bools\dsp \bools \por \bools = \bools\dsp \pnot \bools  = \bools.
\end{equation}
\end{statement}

The boolean operations defines an algebraic structure over \bools\ called boolean algebra: \tuple{\bools,[\pand, \por, \pnot], \sdef{0,1}}, which satisfies the following properties:
\begin{flalign}
\forall a,b,c \in \Bset:&&
\begin{array}{c}
\begin{aligned}
a\pand b &= b \pand a \\
a\por b &= b \por a 
\end{aligned}\qquad
\begin{aligned}
a\pand (b\pand c)&= (a \pand b) \pand c\\
a\por (b\por c)&= (a \por b) \por c
\end{aligned}\qquad
\begin{aligned}
a\pand (a\por b)&= a\\
a\por (a\pand b)&= a
\end{aligned}\vspace{5pt}\\
\begin{aligned}
a\pand \pnot a &= 0\\
a\por \pnot a &= 1
\end{aligned} \qquad
\begin{aligned}
a\pand a &= a\\
a\por a &= a
\end{aligned}\qquad
\begin{aligned}
a\pand 1 &= a\\
a\por 0 &= a
\end{aligned}\qquad
\begin{aligned}
a\pand 0 &= 0\\
a\por 1 &= 1
\end{aligned}\qquad
\begin{aligned}
\pnot 1 &= 0\\
\pnot 0 &= 1
\end{aligned}
\end{array}&&
\end{flalign}

\subsection{The Category of Tautologies \tauts\ and of Contradictions \ctrs}

The two subsets of the boolean category, \sdef{0}\  and \sdef {1}\, defines two special categories: the category of contradictions \ctrs\footnote{ArXiv preprocessor couldn't handle this notation properly. The correct version can be found at \url{https://docs.google.com/open?id=0B7JbxpuZS0IsRFotZ2VuU0lpcTA}.} and the category of tautologies \tauts:
\begin{edefinition}
\ctrs \eqd \sdef{0},
\end{edefinition}
\begin{edefinition}
\tauts \eqd \sdef{1}.
\end{edefinition}

\begin{statement}
The boolean category is the union of the categories of tautologies and contradictions.An element that is either a tautology or a contradiction is a boolean element.
\begin{equation}
\bools = \tauts \un \ctrs.
\end{equation}
\end{statement}

\begin{statement}
The negation of a tautology is a contradiction. The negation of a contradiction is a tautology.
\begin{equation}
\pnot \tauts = \ctrs\dsp \pnot \ctrs = \tauts.
\end{equation}
\end{statement}

\begin{statement}
The conjunction or the disjunction of two tautologies is also a tautology:
\begin{equation}
\tauts \pand \tauts = \tauts\dsp \tauts \por \tauts = \tauts.
\end{equation}
\end{statement}

\begin{statement}
The conjunction or the disjunction of two contradictions is also a contradiction:
\begin{equation}
\ctrs \pand \ctrs = \ctrs\dsp \ctrs \por \ctrs = \ctrs.
\end{equation}
\end{statement}

\begin{statement}
The conjunction of a tautology and a contradiction is a contradiction. The disjunction of a tautology and a contradiction is a tautology.
\begin{equation}
\tauts \pand \ctrs = \ctrs\dsp \tauts \por \ctrs = \tauts.
\end{equation}
\end{statement}

\subsection{The Category of Propositions \prps}

The category of propositions \prps\ is a generalization of the boolean category that contains all the self-adjoint idempotents elements, and so is defined by:
\begin{edefinition}
\prps \eqd \sdef{a \in \ets : a\s = a,\, a^2 = a}.
\end{edefinition}

\begin{interpretation}
Each member of \prps\ is called a proposition, \prps\ defines the concept of proposition.
\end{interpretation}

\begin{statement}
Tautologies and contradictions are propositions. Boolean elements are propositions.
\begin{equation}
\tauts \subset \prps\dsp \ctrs \subset \prps\dsp \bools \subset \prps.
\end{equation}
\end{statement}

\begin{statement}
The negation of a proposition is a proposition.
\begin{equation}
\pnot \prps = \prps.
\end{equation}
\end{statement}

\begin{statement}
The conjunction or the disjunction of two propositions is a proposition if, and only if, they are compatible.
\begin{cequation}{ \prp{a}, \prp{b} \in \prps}
\begin{aligned}
\prp{a} \pand \prp{b} \in \prps &\iff \prp{a} \cpt \prp{b}\\
\prp{a} \por\prp{b} \in \prps &\iff \prp{a} \cpt \prp{b}
\end{aligned}
\end{cequation}
\end{statement}

\begin{statement}
The conjunction or the disjunction of two compatible propositions is a proposition.
\begin{equation}
\pand(\prps \cpt \prps) = \prps\dsp \por(\prps \cpt \prps) = \prps.
\end{equation}
\end{statement}

\begin{statement}
The conjunction of a proposition and a tautology is the proposition itself.
\begin{cequation}{ \prp{a} \in \prps}
\prp{a} \pand 1 = a.
\end{cequation}
\end{statement}

\begin{statement}
The disjunction of a proposition and a contradiction is the proposition itself.
\begin{equation}
\forall \prp{a} \in \prps\dsp \prp{a} \por 0 = a.
\end{equation}
\end{statement}

\begin{statement}
The disjunction of a proposition and a tautology is a tautology.
\begin{equation}
\prps \por \tauts = \tauts.
\end{equation}
\end{statement}

\begin{statement}
The conjunction of a proposition and a contradiction is a contradiction.
\begin{equation}
\prps \pand \ctrs = \ctrs.
\end{equation}
\end{statement}

\begin{formalism}
Between an element and a proposition, the invariance relation simplifies to:
\begin{cequation}{ a \in \ets\bsp \prp{b} \in \prps}
a \pleq \prp{b} \iff \prp{b} a \prp{b} = a.
\end{cequation}
\end{formalism}

\begin{formalism}
Between two compatible propositions the invariance relation simplifies even more:
\begin{cequation}{  (\prp{a},\prp{b}) \in \prps\cpt\prps}
\begin{aligned}
\prp{a} \pleq \prp{b} &\iff  \prp{a} \pand\prp{b} = \prp{a}\\
\prp{a} \pleq \prp{b} &\iff  \prp{a} \por\prp{b} = \prp{b}.
\end{aligned}
\end{cequation}
\end{formalism}

\begin{interpretation}
When two propositions, \prp{a}, \prp{b}, satisfies $\prp{a} \pleq \prp{b}$, we say that [\prp{a} affirms \prp{b}], or that [\prp{b} follows from \prp{a}].
\end{interpretation}

\begin{statement}
A proposition affirms every proposition that follows from it.
\begin{cequation}{ \prp{a}, \prp{p} \in \prps}
\prp{b} \pgeq  \prp{a} \then \prp{a} \pleq  \prp{b}
\end{cequation}
\end{statement}

\begin{interpretation}
When $\prp{a} \pleq \pnot \prp{b}$ holds we say that [\prp{a} negates \prp{b}].
\end{interpretation}

\begin{statement}
When \prp{a}\ and \prp{b}\ are compatible propositions, the proposition $\prp{a} \pand \prp{b}$ affirms both \prp{a}\ and \prp{b}.
\begin{cequation}{ (\prp{a},\prp{b}) \in \prps\cpt\prps}
\begin{aligned}
 \quad \prp{a} \pand \prp{b} \pleq \prp{a},\\
\prp{a} \pand \prp{b} \pleq \prp{b}.
\end{aligned}
\end{cequation}
\end{statement}

\begin{interpretation}
When  $\prp{a} \ctrd \prp{b}$, we say that [\prp{a}\ and \prp{b}\ are contradictory], or that [\prp{a} contradicts \prp{b}].
\end{interpretation}

\begin{statement}
The conjunction of two contradictory propositions is a contradiction.
\begin{equation}
\pand(\prps \ctrd \prps) = \ctrs.
\end{equation}
\end{statement}

\begin{statement}
A tautology follows from any proposition.
\begin{cequation}{ \prp{a} \in \prps}
\prp{a} \pleq 1.
\end{cequation}
\end{statement}

\begin{statement}
A contradiction affirms any proposition.
\begin{cequation}{ \prp{a} \in \prps}
0 \pleq \prp{a}.
\end{cequation}
\end{statement}

\begin{statement}
Every proposition follows form itself. Every proposition affirms itself.
\begin{cequation}{ \prp{a} \in \prps}
\prp{a} \pleq \prp{a}.
\end{cequation}
\end{statement}

\begin{statement}
If a proposition \prp{a}\ affirms \prp{b}, and \prp{b}\ affirms \prp{c}, then \prp{a}\ affirms \prp{c}. If a proposition \prp{a} follows from \prp{b}, and \prp{b} follows from \prp{c}, then \prp{a} follows \prp{c}.
\begin{cequation}{ \prp{a}, \prp{b}, \prp{c} \in \prps}
\begin{aligned}
\prp{a} \pleq \prp{b} \tand  \prp{b} \pleq \prp{c} &\then  \prp{a} \pleq \prp{c},\\
\prp{a} \pgeq \prp{b} \tand  \prp{b} \pgeq \prp{c} &\then  \prp{a} \pgeq \prp{c}.
\end{aligned}
\end{cequation}
\end{statement}

\begin{statement}
If a proposition \prp{a}\ affirms \prp{b}, and \prp{b} affirms \prp{a}, then \prp{a}\ and \prp{b}\ are the same proposition.
\begin{cequation}{ \prp{a}, \prp{b} \in \prps}
\prp{a} \pleq \prp{b} \tand \prp{b} \pleq \prp{a} \then \prp{a} = \prp{b}.
\end{cequation}
\end{statement}

\begin{statement}
Two contradictory propositions negates each other.
\begin{cequation}{ (\prp{a},\prp{b}) \in\prps\ctrd\prps}
\begin{aligned}
\prp{a} &\pleq \pnot \prp{b},\\
 \prp{b} &\pleq \pnot \prp{a}.
\end{aligned}
\end{cequation}

\end{statement}

\section{Bras and Kets}

\subsection{The Categories of Bras \bras\ and Kets \kets}

There are two special categories which are dual vector spaces, the category of bras \bras\ and the category of kets \kets, which are connected by the property that the adjoint of a ket is a bra and the adjoint of a bra is a ket:
\begin{equation}
\kets\s = \bras\dsp \bras\s = \kets.
\end{equation}

We use the following notation: \ket{\;\cdot\;}\ for kets, and \bra{\;\cdot\;}\ for bras, with the property that $\ket{\;\cdot\;}\s = \bra{\;\cdot\;}$ and $\bra{\;\cdot\;}\s = \ket{\;\cdot\;}$.

These two categories satisfies the axioms:
\begin{equation}
\begin{array}{c}
\kets \subset \ets,\\
\kets+\kets = \kets\dsp \Cset \kets = \kets\dsp \kets_* \bras_* \its \bras_*\kets_* = 
\emptyset,\\
\bras\kets=\Cset,
\end{array}
\end{equation}\nobreak\vspace{-20pt}\nobreak
\begin{cequation}{\ket{\psi} \in \kets}
\braket{\psi|\psi}  \geq 0\csp \braket{\psi|\psi} = 0 \iff \ket{\psi}=0.
\end{cequation}
\vspace{15pt}
\begin{statement}
The sum of two kets (bras) is a ket (bra).
\begin{equation}
\kets+\kets = \kets\dsp \bras + \bras = \bras.
\end{equation}
\end{statement}
\begin{statement}
The product of a complex number and a ket (bra) is a ket (bra).
\begin{equation}
\Cset \kets = \kets\dsp \Cset \bras = \bras.
\end{equation}
\end{statement}
\begin{statement}
The product of a bra and a ket is a complex number.
\begin{equation}
\bras\kets=\Cset.
\end{equation}
\end{statement}
\begin{statement}
The product of a non-null ket and a non-null bra is not a complex number. The multiplication of bras and kets is necessarily non commutative for non-null bras and kets.
\begin{equation}
\kets_* \bras_* \its \bras_*\kets_* = \emptyset.
\end{equation}
\end{statement}
\begin{statement}
Minus a ket (bra) is a ket (bra).
\begin{equation}
-\kets=\kets\dsp -\bras=\bras.
\end{equation}
\end{statement}

The categories of bras and kets are vector spaces over the complex numbers:

\noindent$\forall \ket{\psi}\asp\ket{\phi}\asp\ket{\varphi} \in \kets\bsp \alpha\asp \beta \in \Cset: $\vspace{-\baselineskip}
\twocolumneq{
\begin{align}
 \ket{\psi} + \ket{\phi} &= \ket{\phi} + \ket{\psi}\\
(\ket{\psi} + \ket{\phi}) + \ket{\varphi} &= \ket{\psi} + (\ket{\phi} + \ket{\varphi})\\
\ket{\psi} + 0 &= \ket{\psi}\\
\ket{\psi} + (-\ket{\psi}) &= 0
\end{align}
}
{
\begin{align}
\alpha(\ket{\psi} + \ket{\phi}) &= \alpha\ket{\psi} + \alpha\ket{\phi}\\
(\alpha + \beta)\ket{\psi} &= \alpha\ket{\psi} + \beta\ket{\psi}
\eqspace
\alpha(\beta\ket{\psi}) &=(\alpha\beta)\ket{\psi}\\
1\ket{\psi} &= \ket{\psi}
\end{align}
}

\subsubsection{The $\braket{\;\cdot\;|\;\cdot\;}$-Inner Product }

\begin{formalism}
These vector spaces are also inner product vector spaces, that is, there is an inner product operation $\braket{\;\cdot\;|\;\cdot\;}$ defined by:
\begin{equation}
\forall  \ket{\psi}, \ket{\phi} \in \kets: \quad \Braket{\bra{\psi}\big|\ket{\phi}} = \Braket{\ket{\psi}\big|\ket{\phi}} = \Braket{\ket{\psi}\big|\bra{\phi}} = \Braket{\ket{\psi}\big|\bra{\phi}} \eqd \bra{\psi}\ket{\phi}.
\end{equation}
\end{formalism}

This operations takes two bras or two kets or a bra and a ket into a complex number. In the special cases in which there is no risk of ambiguity we can use the economical notation:
\begin{equation}
\Braket{\bra{\psi}\big|\ket{\phi}} = \Braket{\ket{\psi}\big|\ket{\phi}} = \Braket{\ket{\psi}\big|\bra{\phi}} = \Braket{\ket{\psi}\big|\bra{\phi}} = \Braket{\psi|\phi}.
\end{equation}

This operation satisfies:

\noindent$\forall \ket{\psi}\asp\ket{\phi}\asp\ket{\varphi} \in \kets\bsp \alpha\asp \beta \in \Cset: $\vspace{-\baselineskip}
\begin{subequations}
\begin{align}
\Braket{\psi|\phi}\s &= \Braket{\phi|\psi}\\
\Braket{\bra{\psi}\big|\alpha\ket{\phi} + \beta\ket{\varphi}} &= \alpha\Braket{\psi|\phi} + \beta\Braket{\psi|\varphi}\\
\Braket{\psi|\psi} &\in \Rset \geq 0
\end{align}
\end{subequations}

\subsection{The Categories of State Bras \Sbras and of State Kets \Skets}

There are two special subcategories of these categories, the category of state kets \Skets\ and the category of state bras \Sbras\ defined by:
\begin{edefinition}
\Skets \eqd \sdef{\ket{\psi} \in \kets : \braket{\psi|\psi}=1},
\end{edefinition}
\begin{edefinition}
\Sbras \eqd \sdef{\bra{\psi} \in \bras : \braket{\psi|\psi}=1}.
\end{edefinition}

\begin{statement}
The adjoint of a state ket is a state bra. The adjoint of a state bra is a state ket.
\begin{equation}
\Skets\s = \Sbras\dsp \Sbras\s=\Skets.
\end{equation}
\end{statement}

\section{Operators}

\subsection{The Category of Liner Operators \ops}

Related with the categories of bras and kets, there is the category of linear operators \ops, or simply category of operators. Each member of \ops\ is called a linear operator, or simply an operator.

For a matter of simplicity we will only give the definition of finite operators, that is, operators that can be written as a finite sum of a ket multiplies by a bra. We define the category of operators by:
\begin{edefinition}
\ops \eqd \sum \kets \bras.
\end{edefinition}

\begin{statement}
The product of an operator and a ket is a ket. The product of a bra and an operator is a bra.
\begin{equation}
\ops \kets = \kets\dsp \bras \ops = \bras.
\end{equation}
\end{statement}

\begin{statement}
The product of a bra and a operator and a ket is a complex number.
\begin{equation}
\bras \ops \kets=\Cset.
\end{equation}
\end{statement}

\begin{statement}
The sum of two operators is an operator. The product o two operators is an operator. The adjoint of an operator is an operator. The product o a complex number and an operator is an operator.
\begin{equation}
\ops + \ops = \ops\dsp \ops \ops = \ops\dsp \ops\s = \ops\dsp \Cset\ops = \ops.
\end{equation}
\end{statement}

\begin{statement}
There is no non null operator that is a complex number.
\begin{equation}
\ops_* \its \Cset_* = \emptyset.
\end{equation}
\end{statement}

The category of operators is a star-algebra over the complex numbers, so it is an associative algebra over \Cset\ and satisfies:

\noindent$\forall A\asp B\asp C \in \ops\bsp \alpha\asp \beta \in \Cset: $\vspace{-\baselineskip}
\twocolumneq{
\begin{align}
 A + B &= B+ A\\
(A+B)+C &= A+(B+C)\\
A+0 &= A\\
A+(-A)&=0\eqspace
A(BC)&=(AB)C\eqspace
A(B+C) &= A B + A C\\
(A+B)C &= A C+ B C\\
1 A &= A
\end{align}
}
{
\begin{align}
\alpha(A+B) &= \alpha A + \alpha B\\
(\alpha + \beta)A &= \alpha A + \beta B\eqspace
\alpha (\beta A) &= (\alpha \beta) A\\
(\alpha A) (\beta B) &= (\alpha \beta) (A B)\eqspace
(\alpha A + \beta B)\s&=\alpha\s A\s + \beta\s B\s\\
(AB)\s &= B\s A\s\\
A\s\s&=A\\ \notag
\end{align}
}

An operator behaves inside the inner product in the following way:
\begin{cequation}{\ket{\psi}, \ket{\phi} \in \kets\bsp  A \in \ops}
\begin{aligned}
\Braket{\bra{\psi} \big| A \ket{\phi}}&= \Braket{\bra{\psi} A\big|  \ket{\phi}},\\
\Braket{\ket{\psi} \big| A \ket{\phi}}&= \Braket{A\s\ket{\psi} \big|  \ket{\phi}}.
\end{aligned}
\end{cequation}

Note that we use the economical notation only when there is no risk of ambiguity, $\braket{A\s\psi|\phi}$ for example, could be seems ambiguous in some cases and should be avoided, although we would usually be able handle this and extract the correct meaning.

\begin{statement}
The product of a ket and a bra is an operator.
\begin{equation}
\kets\bras \subset \ops.
\end{equation}
\end{statement}

\subsubsection{The $\mu$-Function}

\begin{formalism}
We define the special additive $\mu$-function  by the properties:
\begin{subequations}
\begin{align}
\forall A, B \in \ops: \qquad \msr{A+B} &= \msr{A} + \msr{B},\\
\forall a \in \ets\bsp \ket{\psi}, \ket{\phi} \in \kets: \quad \msr{a\ket{\psi}\bra{\phi}}&=\bra{\phi}a\ket{\psi}.
\end{align}
\end{subequations}
\end{formalism}

This function satisfies:
\begin{subequations}
\begin{align}
\forall A, B \in \ops\bsp \alpha \in \Cset: \quad \msr{A} &\in \Cset,\\
\msr{\alpha A} &= \alpha \msr{A},\\
\msr{AB} &= \msr{BA}.
\end{align}
\end{subequations}
Note that we are working with finite operators and some of these properties do not hold in the most general cases. This function satisfies the same properties as the matrix trace.

\subsubsection{The \norm{\,\cdot\,}-Norm}

\begin{formalism}
The $\mu$ function allow us to define the called Hilbert-Schmidt norm:\nobreak
\begin{cequation}{A \in \ops}
\norm{A} \eqd \sqrt{\msr{\displaystyle{A\s} A}}
\end{cequation}
 which satisfies:
\begin{subequations}
\begin{align}
\forall A,B \in \ops:\quad \norm{AB} &\leq \norm{A} \norm{B},\\
\norm{A\s A} &= \norm{A} \norm{A\s}.
\end{align}
\end{subequations}
\end{formalism}

This properties classifies the category of operators as a well known structure called $C\s$-algebra.

\subsection{The Category of Real Operators \rops}

We define the category of real operators as the set of all self-adjoint operators:
\begin{edefinition}
\rops \eqd \sdef{A \in \ops : A\s=A}.
\end{edefinition}

This is basically the set of self-adjoint operators, but given its importance it receive it's own name.

\begin{statement}
The sum of two real operators is a real operator. The product of a real number and real operator is a real operator. The adjoint of a real operator is a real operator.
\begin{equation}
\rops + \rops = \rops\dsp \Rset \rops = \rops\dsp \rops\s=\rops.
\end{equation}
\end{statement}

\begin{statement}
The product of two compatible real operators is a real operator.
\begin{equation}
\cdot(\rops\cpt\rops)=\rops
\end{equation}
\end{statement}

\begin{statement}
The adjoint of a real operator is the real operator itself.
\begin{cequation}{A \in \rops}
A\s = A
\end{cequation}
\end{statement}

\begin{statement}
The $\mu$ function applied to a real operator is a real number.
\begin{equation}
\msr{\rops} = \Rset.
\end{equation}
\end{statement}

\begin{statement}
A complex operator is the sum of a real operator and the product of the imaginary unity and a real operator.
\begin{equation}
\ops = \rops + i \rops.
\end{equation}
\end{statement}


\subsection{The Category of Positive Operators \pops}

\begin{formalism}
We extend the definition of the order relation $\geq$ to the domain of linear operators by
\begin{cequation}{A, B \in \ops}
A \geq B \rld A-B \geq 0,
\end{cequation}
this last relation being the positivity relation defined by
\begin{cequation}{A \in \ops}
A \geq 0 \rld \forall \ket{\psi} \in \kets:\; \bra{\psi}A\ket{\psi} \in \Rset \geq 0.
\end{cequation}
\end{formalism}

With it, we define the category of positive operators \pops:
\begin{edefinition}
\pops \eqd \sdef{A \in \ops : A \geq 0}.
\end{edefinition}

\begin{interpretation}
We say that an operator [$A$ is positive] iff $A \geq 0$
\end{interpretation}

\begin{interpretation}
We say that an operator [$A$ is greater than or equal to $B$] iff $A \geq B$.
\end{interpretation}

\begin{interpretation}
We say that an operator [$A$ is smaller than or equal to $B$] iff $A \leq B$.
\end{interpretation}

\begin{statement}
An operator is positive iff it can be written as the product of the adjoint of an operator with the operator itself.
\begin{cequation}{A \in \ops}
A \geq 0 \iff \exists B \in \ops: A = B\s B.
\end{cequation}
\end{statement}

\begin{statement}
A positive operator is a real operator.
\begin{equation}
\pops \subset \rops.
\end{equation}
\end{statement}

\begin{statement}
The sum of two positive operators is a positive operator.
\begin{equation}\begin{array}{c}
\pops+\pops = \pops\\
\forall A, B \in \ops:\quad A \geq 0 \tand B \geq 0 \then A+B \geq 0
\end{array}
\end{equation}
\end{statement}

\begin{statement}
The multiplication of a positive operator by a positive number is a positive operator.
\begin{equation}
\pops \Rset_{\geq 0} = \pops.
\end{equation}
\end{statement}

\begin{statement}
The adjoint of a positive operator is a positive operator.
\begin{equation}
\pops\s = \pops.
\end{equation}
\end{statement}

\begin{statement}
The product of two compatible positive operators is a positive operator.
\begin{eqnarray}\begin{array}{c}
+(\pops \cpt \pops) = \pops\\
\forall (A, B) \in (\ops \cpt \ops):\quad A \geq 0 \tand B \geq 0 \then AB \geq 0
\end{array}
\end{eqnarray}
\end{statement}

\begin{statement}
If $A$ is a positive operator and $U$ an operator, the product $U\s A U$ is a positive operator.
\begin{cequation}{A \in \pops \bsp U \in \ops}
 \quad U\s A U \geq 0.
\end{cequation}
\end{statement}

\begin{statement}
A projection of a positive operator is a positive operator.
\begin{cequation}{\prp{P} \in \pjs \bsp A \in \pops}
 \quad \rst{A}{\prp{P}} \in \pops.
\end{cequation}
\end{statement}

\begin{statement}
Between two compatible real operators, the order relation $\leq$ is a partial order.

$\forall (A, B, C)\in \pops \cpt \pops \cpt \pops,$\\[-20pt]
\begin{subequations}
\begin{gather}
A \geq A\\
A \geq B \tand B \geq C \then A \geq C\\
A \geq B \tand B \geq A \then A = B
\end{gather}
\end{subequations}

\end{statement}



These strong symmetry between real numbers and real operators justify why do we call them real operators.

\subsection{The Category of Projectors \pjs}

We define the category of projectors \pjs\ as the set of all operators that are also propositions:
\begin{edefinition}
\pjs \eqd \ops \its \prps = \sdef{A \in \rops : A^2=A}.
\end{edefinition}

\begin{statement}
A projector is a proposition.
\begin{equation}
\pjs \subset \prps.
\end{equation}
\end{statement}

\begin{statement}
A projector is a real positive operator.
\begin{equation}
\pjs \subset \rops\dsp \pjs \subset \pops\dsp \pjs \subset \ops.
\end{equation}
\end{statement}

\begin{statement}
The conjunction or disjunction of two compatible projectors is also a projector
\begin{equation}
\pand(\pjs \cpt \pjs) = \pjs\dsp \por(\pjs \cpt \pjs) = \pjs.
\end{equation}
\end{statement}

\begin{statement}
The negation of a projector is not a projector.
\begin{equation}
\pnot \pjs \its \pjs = \emptyset.
\end{equation}
\end{statement}

\begin{statement}
Tautologies or contradictions are not projectors.
\begin{equation}
\tauts \its \pjs = \emptyset\dsp \ctrs \its \pjs = \emptyset.
\end{equation}
\end{statement}

Any projector can be written as a finite sum of products of a state ket with its corresponding state ket, that is:
\begin{equation}
\forall \prp{A} \in \pjs:\quad \prp{A} = \sum_{i=1}^{n} \ket{\phi_i}\bra{\phi_i}\csp \braket{\phi_i|\phi_j}=\delta_{i,j}\csp n \in \Nset,
\end{equation}
and then $\msr{A}=n \in \Nset$.

\begin{interpretation}
For any projector \prp{A}, we call \msr{A}\ the [rank of \prp{A}].
\end{interpretation}

\begin{statement}
The rank of a projector is a natural number.
\begin{equation}
\msr{\pjs}=\Nset.
\end{equation}
\end{statement}

\begin{statement}
If a projector \prp{A} affirms a projector \prp{B}, then the rank of \prp{A} is smaller than or equal to the rank of \prp{B}.
\begin{cequation}{\prp{A}, \prp{B} \in \pjs}
\prp{A} \pleq \prp{B} \then \msr{\prp{A}} \leq \msr{\prp{B}}.
\end{cequation}
\end{statement}

\begin{statement}
The rank of the disjunction of two contradictory projector is the sum of the ranks of each projector.
\begin{cequation}{(\prp{A}, \prp{B}) \in \pjs \ctrd \pjs}
 \msr{\prp{A} \por \prp{b}} = \msr{\prp{A}} + \msr{\prp{A}}.
\end{cequation}
\end{statement}

\begin{statement}
The rank of the disjunction of two compatible projectors \prp{A}\ and \prp{B}\ is the sum of the individual ranks minus the rank of the sum of their conjunction.
\begin{cequation}{(\prp{A}, \prp{B}) \in \pjs \cpt \pjs}
\msr{\prp{A} \por \prp{b}} = \msr{\prp{A}} + \msr{\prp{A}}-\msr{\prp{A}\pand\prp{B}}
\end{cequation}
\end{statement}

\section{Subspaces of Kets and Bras, and sub algebras of Operators}

Any subset $\kets_i$ of kets satisfying:
\begin{equation}
\kets_i \subseteq \kets\dsp \kets_i + \kets_i = \kets_i\dsp \Cset \kets_i = \kets_i,
\end{equation}
is a subspace of the space of kets and so satisfies all the properties of an inner product vector space.

Associated with any subspace $\kets_i$ of kets there is the corresponding subspace of bras $\bras_i=\kets_i\s$:
\begin{equation}
\bras_i \subseteq \bras\dsp \bras_i + \bras_i = \bras_i\dsp \Cset \bras_i = \bras_i.
\end{equation}

Associate with any subspace $\kets_i$ of kets we also have the corresponding operator algebra $\ops_i$:
\begin{equation}
\ops_i = \sum \kets_i\bras_i,
\end{equation}
satisfying:
\begin{eqnarray}\begin{array}{c}
\ops_i + \ops_i = \ops_i\dsp \ops_i \ops_i = \ops_i\dsp \ops_i\s = \ops_i\dsp \Cset\ops_i = \ops_i,\\
\ops_i \kets_i = \kets_i\dsp \bras_i \ops_i = \bras_i.
\end{array}
\end{eqnarray}

\subsection{Subspaces Characterized by Propositions}
\begin{formalism}
Any proposition \prp{p} defines the subspaces of bras \brarst{\prp{p}}, kets \ketrst{\prp{p}}\ and the corresponding operator algebra \oprst{\prp{p}} according to:
\begin{align}
\forall \prp{p} \in \prps:\quad \brarst{\prp{p}} &\eqd \sdef{\bra{\psi} \in \bras : \bra{\psi} \prp{p} = \ket{\psi}},\\
\ketrst{\prp{p}} &\eqd \sdef{\ket{\psi} \in \kets : \prp{p}\ket{\psi}  = \ket{\psi}},\\
\oprst{\prp{p}} &\eqd \sdef{A \in \ops : A \pleq \prp{p}},
\end{align}
and also we define:
\begin{align}
\Sketrst{\prp{p}} &\eqd \Skets \its \ketrst{\prp{p}},\\
\Sbrarst{\prp{p}} &\eqd \Sbras \its \brarst{\prp{p}},\\
\roprst{\prp{p}} & \eqd \rops \its \oprst{\prp{p}}\\
\poprst{\prp{p}} & \eqd \pops \its \oprst{\prp{p}}\\
\pjrst{\prp{p}} &\eqd \pjs \its \oprst{\prp{p}}.
\end{align}
\end{formalism}

\begin{interpretation}
We say that the proposition \prp{p}\ characterizes the set of kets \ketrst{\prp{p}}. The set \ketrst{\prp{p}}\ is the set of kets characterized by \prp{p}.
\end{interpretation}

\begin{statement}
The conjunction or disjunction of compatible projectors characterized by a proposition is also a projector characterized by the same proposition.
\begin{equation}
\pand(\pjrst{\prp{p}} \cpt \pjrst{\prp{p}}) = \pjrst{\prp{p}}\dsp \por(\pjrst{\prp{p}} \cpt \pjrst{\prp{p}}) = \pjrst{\prp{p}}.
\end{equation}
\end{statement}

\begin{formalism}
We extend the definition to allow sets to be define by more than one proposition:
\begin{equation}
\rst{\Xset}{\prp{p},\prp{q},\ldots} \eqd \rst{\Xset}{\prp{p}} \its \rst{\Xset}{\prp{q}} \its \cdots,
\end{equation}
and also by sets of propositions:
\begin{equation}
\rst{\Xset}{\sdef{\prp{p},\prp{q}, \ldots}} \eqd \rst{\Xset}{\prp{p},\prp{q},\ldots}.
\end{equation}
\end{formalism}

\begin{interpretation}
We say that \rst{\Xset}{\prp{p},\prp{q}} is the subset of \Xset\ characterized by \prp{p}\ and by \prp{q}, or simply by \prp{p}\ and \prp{q}.
\end{interpretation}

\begin{statement}
A set characterized by \prp{p}\ and by \prp{q}\ is the same as the one characterized by \prp{q}\ and by \prp{q}.
\begin{cequation}{\prp{p}, \prp{q} \in \prps}
\rst{\Xset}{\prp{p},\prp{q}} = \rst{\Xset}{\prp{q},\prp{p}}.
\end{cequation}
\end{statement}

\begin{statement}
A set characterized by \prp{p}\ and by \prp{p}\ is the same as the one characterized just by \prp{p}.
\begin{cequation}{\prp{p} \in \prps}
\rst{\Xset}{\prp{p},\prp{p}} = \rst{\Xset}{\prp{p}}
\end{cequation}
\end{statement}

\begin{statement}
If \prp{p}\ affirms \prp{q}, then a set characterized by \prp{p}\ is contained in the set characterized by \prp{q}.
\begin{cequation}{\prp{p}, \prp{q} \in \prps}
\prp{p} \pleq \prp{q} \then \rst{\Xset}{\prp{p}} \subseteq \rst{\Xset}{\prp{q}}
\end{cequation}
\end{statement}

\begin{statement}
The set characterized by conjunction of two compatible propositions $\prp{p}\pand \prp{q}$ is the set characterized by \prp{p}\ and by \prp{q}.
\begin{cequation}{(\prp{p},\prp{q}) \in \prps \cpt \prps}
\rst{\Xset}{\prp{p}\pand \prp{q}} =\rst{\Xset}{\prp{p},\prp{q}} = \rst{\Xset}{\prp{p}} \its \rst{\Xset}{\prp{q}}.
\end{cequation}
\end{statement}

\begin{statement}
The set characterized by disjunction of two compatible propositions, $\prp{p}\pand \prp{q}$, is the sum of sets characterized by \prp{p}\ and the set characterized by \prp{q}.
\begin{cequation}{(\prp{p},\prp{q}) \in \prps \cpt \prps}
\rst{\Xset}{\prp{p}\pand \prp{q}} = \rst{\Xset}{\prp{p}} + \rst{\Xset}{\prp{q}}.
\end{cequation}
\end{statement}

\begin{statement}
If \prp{p} contradicts \prp{q}, then the set characterized by \prp{p}\ and \prp{q} is a subset of the category of contradictions.
\begin{cequation}{(\prp{p},\prp{q}) \in \prps \ctrd \prps}
\rst{\Xset}{\prp{p}, \prp{q}} \subseteq \ctrs.
\end{cequation}
\end{statement}

The dimension of the subspace of bras or kets characterized by a projector is the rank of the projector. If
\[
\prp{P} =  \sum_{i=1}^{n} \ket{\psi_i}\bra{\psi_i}\dsp \braket{\psi_i|\psi_j} = \delta_{i,j},
\]
\nobreak then the subspace \ketrst{\prp{P}}\ posses the set \sdef{\ket{\psi_i}: i=1,\cdots,n} as one of its orthogonal basis, that is,
\[
\ketrst{\prp{P}} = \sum_{i=1}^n \Cset \ket{\psi_i}.
\]
\nobreak In this case, $n$ is the dimension of this space.

\begin{statement}
A sub-algebra \oprst{\prp{P}} characterized by a projector \prp{P}\ is a unital algebra with \prp{P}\ as its multiplicative identity element.
\begin{cequation}{\prp{P} \in \pjs\bsp A \in \oprst{\prp{P}}}
 A \prp{P} = \prp{P} A = A.
\end{cequation}
\end{statement}

\begin{formalism}
We define a class of new operations on propositions called complement of type \prp{p}, or simply \prp{p}-complement, where \prp{p} is proposition, by:
\begin{cequation}{\prp{p} \in \prps\bsp a \in \ets}
\pcmp{\prp{p}}a \eqd \prp{p} - a.
\end{cequation}
\end{formalism}

\begin{statement}
The \prp{P}-complement of a projector characterized by \prp{P} is also a projector characterized by \prp{P}.
\begin{equation}
\pcmp{\prp{P}} \pjrst{\prp{p}} = \pjrst{\prp{p}}.
\end{equation}
\end{statement}

\begin{statement}
If \prp{A} is a projector characterized by \prp{P}, the conjunction of \prp{A} and its \prp{P}-complement is a contradiction, and the disjunction of \prp{A} and its \prp{P}-complement is the projector \prp{P}, the  \prp{P}-complement of the \prp{P}-complement of \prp{A}\ is \prp{A}\ itself.
\begin{cequation}{\prp{p} \in \pjs\bsp  \prp{A} \in \pjrst{\prp{P}}}
\begin{aligned}
\prp{A} \pand \pcmp{\prp{P}} &= 0\\
\prp{A} \por \pcmp{\prp{P}} &= \prp{P}\\
\pcmp{\prp{P}}\pcmp{\prp{P}}\prp{A} &= \prp{A}
\end{aligned}
\end{cequation}
\end{statement}

\subsection{Projection}

\begin{formalism}
We define the projection of an operator $A$ by a projector \prp{P} by
\begin{cequation}{\prp{P} \in \pjs\bsp a \in \ops}
\rst{a}{\prp{P}} \eqd \prp{P}a\prp{P}.
\end{cequation}
\end{formalism}

\begin{interpretation}
The projection of an operator $A$ by a projector \prp{P}\ is denoted by  \rst{a}{\prp{P}}.
\end{interpretation}

\begin{statement}
The projection of an operator belongs to the algebra of operators characterized by the projector.
\begin{cequation}{\prp{P} \in \pjs\bsp A \in \ops}
\rst{A}{\prp{P}} \in \oprst{\prp{P}}.
\end{cequation}

\end{statement}

\begin{interpretation}
We say that an operator [$A$ is invariant under projection by \prp{P}] iff $A \pleq \prp{P}$.
\end{interpretation}

\subsection{The Category of Elementary Projectors or State of Affairs \Spjs}

\begin{formalism}
We define an special operation \stp{\,\cdot\,} on bras and kets, by
\begin{cequation}{\ket{\psi} \in \kets}
\stp{\ket{\psi}}=\stp{\bra{\psi}} \eqd \ket{\psi}\bra{\psi}.
\end{cequation}
\end{formalism}

When there is no risk of ambiguity we can use the economical notation:
\begin{equation}
\stp{\ket{\psi}}=\stp{\bra{\psi}} = \stp{\psi}
\end{equation}

\begin{statement}
The result of \stp{\,\cdot\,}-operation is always a real operator.
\begin{equation}
\stp{\bras} \subset \rops\dsp
\stp{\kets} \subset \rops.
\end{equation}
\end{statement}

\begin{statement}
When acting on state bras or kets the result the  \stp{\,\cdot\,}-operation is a special projector, a projector with rank one:
\begin{eqnarray}
\stp{\Sbras} \subset \pjs\dsp \msr{\stp{\Sbras}}=\tauts,\\
\stp{\Skets} \subset \pjs\dsp \msr{\stp{\Skets}}=\tauts.
\end{eqnarray}
\end{statement}

Now we define the category of elementary projectors \Spjs:
\begin{edefinition}
\Spjs \eqd \sdef{\prp{A} \in \pjs \;\Big|\; \msr{\prp{A}} = 1}.
\end{edefinition}

\begin{statement}
An elementary projector is a projector with rank one.
\begin{equation}
\msr{\Spjs}=\tauts.
\end{equation}
\end{statement}

\begin{statement}
The result of \stp{\,\cdot\,}-operation on bras or on kets is an elementary projector.
\begin{equation}
\stp{\Skets} = \Spjs\dsp \stp{\Sbras} = \Spjs.
\end{equation}
\end{statement}

\begin{interpretation}
An elementary projector is also called a state o affairs or a state of things.
\end{interpretation}
The reason of it will be shown to be because they can be used to represent the configuration of objects or systems.

\begin{formalism}
We define a special binary relation between two positive operators by
\begin{cequation}{a,B \in \pops}
a \idsc B \rld B a = \alpha a\tand \alpha \in \Rset > 0.
\end{cequation}
\end{formalism}

\begin{formalism}
Between a state of affairs and a positive operator this relation satisfies:
\begin{equation}
\stp{\psi} \idsc A \then \forall \prp{p} \in \prps:\; \prp{p} \pgeq A \then \prp{p} \pgeq \stp{\psi}.
\end{equation}
\end{formalism}

\begin{interpretation}
When a positive operator $A$ satisfies $\stp{\psi} \idsc A$, where $\stp{\psi}$ is an elementary projector, we say that [$A$ describes $\stp{\psi}$], or that, [\stp{\psi}\ is described by $A$].
\end{interpretation}

\chapter{The Nature of Reality}

\section{Objects, Universe and Reality}

Objects are what exists, what propositions are about. Anything we can identify in the physical world is an object, an atom, a molecule, an apple, a planet, a galaxy. For now it is a primitive notion.

Every object in an instance of a certain type. The type of an object is an unchanging part of it. It is what defines the answer to `What is this object?'. Is also a common factor of all the objects of a certain kind, for example, for example, what is common to all the cell phones of a certain model? They have the same type.

This notion of type is also very close to the philosophical notion of form present in Plato and Wittgenstein\cite{Wittgenstein}, and also to the concept of type in computer science.

The type of an object in unchangeable. What change is it's state. Not every object has a defined state. This is related to a phenomenon known as entanglement.

There is special case of object that have a defined state no matter how is the state. In the language of quantum theory, these states are separable, that is, not entangled, and evolves trough unitary transformations.

Such an object defines \emph{an universe.}

An isolated not entangled system is a universe.

The totality of all physical objects is \emph{the Physical Universe}, or simply \emph{the Universe}. We believe that the Physical Universe is an universe.

If we are talking about physical objects, the same laws that are valid for the (whole) Universe is valid for a particular universe. Thats why we call them both universes. Without this principle physics would not make any sense. Most of physics is based on the idea that the laws that are valid on the part are valid on whole.

The laws of physics apply not to physical objects, but to physical universes. In general, no conservation law applies to a single object, only to isolated objects.

Any universe posses a state. The state of a universe is a state of affairs, or, an elementary projector.

Any universe posses a type. The type of an universe is a projector.

The type of the universe defines in which states it can be. The state of an universe is a state of affairs characterized by the type of the universe.

The universe is not a concept, but an entity. The universe refers to a defined entity, the universe that we are talking about. We can only talk about a single universe at a time. To talk about more we must give them names, but then they became different objects contained in a larger universe.

\emph{Our language is tied to the universe.}

\textbf{The projector \unity\ is the type of the universe. $\unity \in \pjs$.}
\begin{interpretation}
 \unity\ denotes the [type of the universe].
\end{interpretation}
\begin{statement}
The type of the universe is a projector.
\begin{equation}
\unity \in \pjs.
\end{equation}
\end{statement}
\textbf{The state of affairs \rlty\ is the state of the universe. $\rlty \in \pjsrt$.}
\begin{interpretation} 
\rlty\ denotes [reality].
\end{interpretation}
Reality is the actual state of the universe, not as it could eventually be or we could imagine.

The set of elementary projectors characterized by the type of the universe defines the concept of possible state of affairs \soas:
\begin{edefinition}
\soas \eqd \pjsrt
\end{edefinition}
\begin{statement}
Reality is a possible state of affairs.
\begin{equation}
\rlty \in \soas.
\end{equation}
\end{statement}
\begin{interpretation}
An element is [a possible state of affairs] iff it belongs to \soas.
\end{interpretation}
\begin{interpretation}
A state of affairs is possible iff it belongs to \soas.
\end{interpretation}
Given the type of the universe, all the possible states of affairs are given.
\begin{interpretation}
When $x$ is a positive operator and a proposition \prp{p}\ satisfies $\prp{p} \pgeq x$ we say that [\prp{p}\ depicts $x$].
\end{interpretation}
\begin{interpretation}
When a proposition \prp{p}\ satisfies $\prp{p} \pgeq \rlty$, we say that [\prp{p}\ depicts reality].
\end{interpretation}
\begin{statement}
If \prp{p} depicts reality, then reality is a state of affairs characterized by \prp{p}.
\begin{cequation}{\prp{p} \in \prps}
\prp{p} \pgeq \rlty \iff \rlty \in \Spjsrst{\prp{p}}.
\end{cequation}
\end{statement}

\section{Truth and Falsity}

\subsection{The Concept of Truth or Fact \trus}

We define the concept of truth or fact \trus\ by:
\begin{edefinition}
\trus \eqd \sdef{\prp{p} \in \prps : \prp{p} \pgeq \rlty}.
\end{edefinition}

\begin{interpretation}
When a proposition depicts reality we say that it is true. A proposition [\prp{p}\ is true] iff $\prp{p} \pgeq \rlty$ holds.
\end{interpretation}

\begin{interpretation}
A proposition [\prp{p}\ is a truth or a fact] iff $\prp{p} \in \trus$ holds. A proposition is a truth iff it belongs to the concept of truth.
\end{interpretation}

\begin{statement}
A fact or a truth is a true proposition. Facts depicts reality. Reality affirms the facts. Facts follow from reality. Reality is the state of affairs characterized by the facts.
\begin{cequation}{\prp{p} \in \trus}
\prp{p} \pgeq \rlty.
\end{cequation}
\end{statement}

Truth and fact are synonyms. The concept of truth and the concept of fact are the same.

\begin{statement}
A tautology is always a truth. A contradiction is never a truth.
\begin{equation}
\tauts \subset \trus\dsp \ctrs \nsubseteq \trus.
\end{equation}
\end{statement}

\begin{statement}
The conjunction or disjunction of compatible facts or truths is also a fact or truth.
\begin{equation}
\pand(\trus \cpt \trus) = \trus\dsp \por(\trus \cpt \trus) = \trus.
\end{equation}
\end{statement}

\begin{statement}
If \prp{a}\ affirms \prp{b}\ and \prp{a} is true, then \prp{b} is also true. If \prp{b}\ follows from \prp{a}\ and \prp{a}\ is true, then \prp{b} is also true.
\begin{cequation}{\prp{a}, \prp{b} \in \prps}
a \pleq b \tand a \in \trus \then b \in \trus.
\end{cequation}
\end{statement}

\subsection{The Concept of Falsity \falss}

\begin{interpretation}
When the negation of a proposition depicts reality we say that it is false. A proposition [\prp{p}\ is false] iff $\pnot \prp{p} \pgeq \rlty$ holds.
\end{interpretation}

We define the concept of falsity \falss\ by:
\begin{edefinition}
\falss \eqd \sdef{\prp{p} \in \prps : \pnot \prp{p} \pgeq \rlty}.
\end{edefinition}

\begin{interpretation}
A proposition [\prp{p}\ is a falsity] iff $\prp{p} \in \falss$ holds.  A proposition is a falsity if it belong to the concept of falsity.
\end{interpretation}

\begin{statement}
Reality negates all the falsities.
\begin{cequation}{\prp{p} \in \falss}
\rlty \pleq \pnot \prp{p}.
\end{cequation}
\end{statement}

\begin{statement}
A contradiction is always a falsity. A tautology is never a falsity. 
\begin{equation}
\ctrs \subset \falss\dsp \tauts \nsubseteq \falss.
\end{equation}
\end{statement}

\begin{statement}
No proposition is both a truth and a falsity. No proposition is a true and false at the same time. \emph{This is called the law of non contradiction.}
\begin{equation}
\trus \its \falss = \emptyset.
\end{equation}
\end{statement}

\begin{statement}
If \prp{a}\ negates \prp{b}\ and \prp{a}\ is true, then \prp{b}\ is false. If \prp{a}\ negates \prp{b}\ and \prp{a}\ is a fact (or truth), then \prp{b}\ is falsity.
\begin{cequation}{\prp{a}, \prp{b} \in \prps}
a \pleq \pnot b \tand a \in \trus \then b \in \falss.
\end{cequation}
\end{statement}

\begin{statement}
If \prp{a}\ and \prp{b}\ are contradictory propositions, then, if \prp{a}\ is true, \prp{b}\ is false and if \prp{b}\ is true, \prp{a}\ is false.
\begin{cequation}{\prp{a},\prp{b} \in \prps \ctrd \prps}
\big(\prp{a} \in \trus \iff \prp{b} \in \falss\big) \tand \big(\prp{b} \in \trus \iff \prp{a} \in \falss\big).
\end{cequation}
\end{statement}

\begin{statement}
There are no two contradictory truths. There are no two contradictory falsities.
\begin{equation}
\trus \ctrd \trus = \emptyset\dsp \falss \ctrd \falss = \emptyset.
\end{equation}
\end{statement}

\begin{statement}
The conjunction or disjunction of compatible falsities is also a falsity.
\begin{equation}
\pand(\falss \cpt \falss) = \falss\dsp \por(\falss \cpt \falss) = \falss.
\end{equation}
\end{statement}

\begin{statement}
The negation of a truth is a falsity. The negation of a falsity is a truth.
\begin{equation}
\pnot \trus = \falss\dsp \pnot \falss = \trus.
\end{equation}
\end{statement}

\subsection{The Concept of Classical Proposition \Cprps}

We define the concept of classical proposition \Cprps\ as the set of all proposition that are compatible with reality:
\begin{edefinition}
\Cprps \eqd \sdef{\prp{p} \in \prps : \prp{p} \cpt \rlty}
\end{edefinition}

\begin{interpretation}
We say that a proposition is classical iff it is compatible with reality. A proposition [\prp{p}\ is classical] iff $\prp{p} \in \Cprps$ holds.
\end{interpretation}

\begin{statement}
A proposition is classical iff it is either true of false. The concept of classical proposition is the union of the concepts of truth and falsity.
\begin{equation}
\Cprps = \trus \un \falss.
\end{equation}
\end{statement}

The principle of excluded middle is one of the laws of thought of classical logic. Here it appear as a special property that characterizes the classical propositions. This justifies why do we call these propositions classical propositions.

\begin{statement}
A classical proposition is either true of false. \emph{Classical propositions satisfies the principle of the excluded middle.}
\begin{cequation}{\prp{p} \in \Cprps}
\prp{p} \in \trus \tor  \prp{p} \in \falss.
\end{cequation}
\end{statement}

\begin{statement}
Facts and falsities are classical propositions.
\begin{equation}
\trus \subset \Cprps\dsp \falss \subset \Cprps.
\end{equation}
\end{statement}

\begin{statement}
Not every proposition is a classical proposition. The principle of excluded middle is not valid in the general case.
\begin{equation}
\prps \neq \Cprps\dsp \Cprps \subset \prps.
\end{equation}
\end{statement}

The principle of excluded middle says that a proposition is either true or false. The non validity of this principle is one of the most distinguished properties of the quantum world. The laws of classical logic are not valid in the quantum world. Classical logic is just a particular case. Many of our intuitions are just classical intuitions and so are not valid within the quantum world.

When all the propositions we are using are classical, we say that we are in a classical world. So classical physics is a particular case in which we are in a classical world. In fact we are never in a classical reality because the reality itself is not classical or quantum, what happens is that we can look to reality using `classical lens', that is, all the proposition we are using are classical. The distinction is on the set of propositions we are using. Our measuring apparatus and our senses defines the set of propositions we are using.

\begin{statement}
The conjunction of a truth and a compatible falsity is a falsity. The disjunction of a truth and a compatible falsity is a truth.
\begin{equation}
\pand(\trus \cpt \falss) = \falss\dsp \por(\trus \cpt \falss) = \trus.
\end{equation}
\end{statement}

\begin{statement}
The conjunction or the disjunction of compatible classical propositions is a classical proposition. The negation of a classical proposition is a classical proposition.
\begin{equation}
\pand(\Cprps \cpt \Cprps) = \Cprps\dsp \por(\Cprps \cpt \Cprps) = \Cprps\dsp \pnot\Cprps  = \Cprps.
\end{equation}
\end{statement}

\section{Necessity and Possibility}

\subsection{The Concept of Necessary Truth \Ntrus}

We define the concept of necessary truth as the set of propositions that depicts the type of the universe:
\begin{edefinition}
\Ntrus \eqd \sdef{\prp{p} \in \prps : \prp{p} \pgeq \unity}.
\end{edefinition}

\begin{interpretation}
A proposition is necessary or necessarily true iff it is a necessary truth. A proposition [\prp{p}\ is necessary] iff $\prp{p} \pgeq \unity$ holds.
\end{interpretation}

\begin{statement}
Every necessary truth is a truth. Every necessary proposition is true.
\begin{equation}
\Ntrus \subseteq \trus.
\end{equation}
\end{statement}

\subsection{The Concept of Necessary Falsity \Nfalss}

We define the concept of necessary falsity \Nfalss\ to be the negation of the concept of necessary truth, so:
\begin{edefinition}
\Nfalss \eqd \sdef{\prp{p} \in \prps : \pnot\prp{p} \pgeq \unity}.
\end{edefinition}

\begin{interpretation}
A proposition is impossible or necessarily false iff it is a necessary falsity. A proposition [\prp{p}\ is impossible or necessarily false] iff $\pnot \prp{p} \pgeq \unity$ holds.
\end{interpretation}
\begin{statement}
A necessary falsity is always a falsity. An impossible proposition is always false.
\begin{equation}
\Nfalss \subseteq \falss.
\end{equation}
\end{statement}
\begin{statement}
The negation of a necessary truth is a necessary falsity. The negation of a necessary falsity is a necessary truth.
\begin{equation}
\pnot \Ntrus = \Nfalss\dsp \pnot \Nfalss = \Ntrus.
\end{equation}
\end{statement}
\begin{statement}
A tautology is always a necessary truth. A contradiction is always a necessary falsity. 
\begin{equation}
\tauts \subset \Ntrus\dsp \ctrs \subset \Nfalss.
\end{equation}
\end{statement}
\begin{statement}
The conjunction of the disjunction of two compatible necessary truths is also a necessary truth.
\begin{equation}
\pand(\Ntrus \cpt \Ntrus) = \Ntrus\dsp \por(\Ntrus \cpt \Ntrus) = \Ntrus.
\end{equation}
\end{statement}
\begin{statement}
The conjunction of the disjunction of two compatible necessary falsities is also a necessary falsity.
\begin{equation}
\pand(\Nfalss \cpt \Nfalss) = \Nfalss\dsp \por(\Nfalss \cpt \Nfalss) = \Nfalss.
\end{equation}
\end{statement}
\begin{statement}
If \prp{p}\ affirms \prp{q} and \prp{p}\ is a necessary truth, then \prp{q}\ is also a necessary truth.
\begin{cequation}{\prp{p}, \prp{q} \in \prps}
\prp{p} \pleq \prp{q} \tand \prp{p} \in \Ntrus \then \prp{q} \in \Ntrus.
\end{cequation}
\end{statement}
\begin{statement}
If \prp{p}\ negates \prp{q} and \prp{p}\ is a necessary truth, then \prp{q}\ is  a necessary falsity.
\begin{cequation}{\prp{p}, \prp{q} \in \prps}
\prp{p} \pleq \pnot\prp{q} \tand \prp{p} \in \Ntrus \then \prp{q} \in \Nfalss.
\end{cequation}
\end{statement}

\subsection{The Concept of Possible Truth \Ptrus}

We define the concept of possible truth or simply possible proposition \Ptrus\ as the propositions that are not impossible or not necessarily false, that is:
\begin{edefinition}
\Ptrus \eqd \prps \backslash \Nfalss.
\end{edefinition}

\begin{interpretation}
 A proposition [\prp{p}\ is possible or possibly true] iff $\pnot\prp{p} \not\pgeq \unity$ holds.
\end{interpretation}

\subsection{The Concept of Possible Falsity \Pfalss}

We define the concept of possible falsity \Pfalss\ as the propositions that are not necessary or not necessarily true, that is:
\begin{edefinition}
\Pfalss \eqd \prps \backslash \Ntrus.
\end{edefinition}

\begin{interpretation}
 A proposition [\prp{p}\ is  possibly false] iff $\prp{p} \not\pgeq \unity$ holds.
\end{interpretation}

A proposition that is a possible truth and a possible falsity is what we call a contingent proposition.

\section{Knowledge}

\subsection{Observer}
An observer of a certain universe is something or someone that posses knowledge about the reality or the state of that universe.

An observer posses a state of knowledge which describes reality.

A state of knowledge is a positive operator characterized by the type of the universe. The set \poprt\ defines the concept of state of knowledge.

\subsection{The Concept of Possible State of Knowledge \soks}

We define the concept of possible state of knowledge \soks\ as the set of states of knowledge that describes reality:
\begin{edefinition} 
\soks \eqd \sdef{\rho \in \poprt                                                         : \rlty \idsc \rho}.
\end{edefinition}

Associated with an observer $i$ there is a possible state of knowledge $\state_i$. The state knowledge of an observer contains all the relevant information about the observer.

\begin{interpretation}
We say that $\state_i$ is [the state of knowledge of the observer $i$].
\end{interpretation}
\begin{statement}
The state of knowledge of an observer is a positive operator that describes reality.
\begin{cequation}{\state \in \soks}
\rlty \idsc \state\csp \state \in \pops.
\end{cequation}
\end{statement}

\subsection{The Concept of Known Truth \Ktrus{i}}

We define the concept of known truth or known fact of the observer $i$ by:
\begin{edefinition}
\Ktrus{i} \eqd \sdef{\prp{p} \in \prps : \prp{p} \pgeq \state_i}.
\end{edefinition}
\begin{interpretation}
We say that the observer $i$ knows that the proposition \prp{p} is true iff $\prp{p} \pgeq \state_i$ holds.
\end{interpretation}
\begin{interpretation}
When $\prp{p} \in \Ktrus{i}$ holds, we say that [\prp{p}\ is a known truth or known fact of the observer $i$].
\end{interpretation}
\begin{statement}
A known truth (fact) is always a truth (fact).
\begin{cequation}{i}
\Ktrus{i} \subseteq \trus.
\end{cequation}
\end{statement}
\begin{statement}
If \prp{p}\ affirms \prp{q} and \prp{p}\ is a a known truth, then \prp{q}\ is also a known truth.
\begin{cequation}{\prp{p}, \prp{q} \in \prps}
\prp{p} \pleq \prp{q} \tand \prp{p} \in \Ktrus{i} \then \prp{q} \in \Ktrus{i}.
\end{cequation}
\end{statement}
\begin{statement}
If \prp{p}\ negates \prp{q} and \prp{p}\ is a a known truth, then \prp{q}\ is also a known falsity.
\begin{cequation}{\prp{p}, \prp{q} \in \prps}
\prp{p} \pleq \pnot\prp{q} \tand \prp{p} \in \Ktrus{i} \then \prp{q} \in \Kfalss{i}.
\end{cequation}
\end{statement}

\subsection{The Concept of Known Falsity \Kfalss{i}}

And we define the concept of known falsity of the observer $i$ by:
\begin{edefinition}
\Kfalss{i} \eqd \sdef{\prp{p} \in \prps : \pnot\prp{p} \pgeq \state_i}.
\end{edefinition}
\begin{interpretation}
We say that the observer [$i$ knows that the proposition \prp{p} is false] iff $\pnot\prp{p} \pgeq \state_i$ holds.
\end{interpretation}
\begin{interpretation}
When $\prp{p} \in \Kfalss{i}$, we say that [\prp{p}\ is a known falsity of the observer $i$].
\end{interpretation}
\begin{statement}
 A known falsity is always a falsity.
\begin{cequation}{i}
\Kfalss{i} \subseteq \falss.
\end{cequation}
\end{statement}
\begin{statement}
A necessary truth is always a known truth. A necessary falsity is always a known falsity. If a proposition is necessary, it is known by every observer.
\begin{cequation}{i}
\Ntrus \subseteq \Ktrus{i}\dsp \Nfalss \subseteq \Kfalss{i}.
\end{cequation}
\end{statement}
\begin{statement}
A tautology is a known fact of every observer. A contradiction is a known falsity of every observer.
\begin{cequation}{i}
\tauts \subset \Ktrus{i}\dsp \ctrs \subset \Kfalss{i}.
\end{cequation}
\end{statement}
\begin{statement}
The negation of a known truth is a known falsity. The negation of a known falsity is a known truth.
\begin{cequation}{i}
\pnot \Ktrus{i} = \Kfalss{i}\dsp \pnot \Kfalss{i} = \Ktrus{i}.
\end{cequation}
\end{statement}
\begin{statement}
The conjunction or the disjunction of compatible known truths is also a known truth.
\begin{cequation}{i}
\pand(\Ktrus{i}\cpt\Ktrus{i}) = \Ktrus{i}\dsp \por(\Ktrus{i}\cpt\Ktrus{i}) = \Ktrus{i}
\end{cequation}
\end{statement}
\begin{statement}
The conjunction or the disjunction of compatible known falsities is also a known falsity.
\begin{cequation}{i}
 \pand(\Kfalss{i}\cpt\Kfalss{i}) = \Kfalss{i}\dsp \por(\Kfalss{i}\cpt\Kfalss{i}) = \Kfalss{i}.
\end{cequation}
\end{statement}
\begin{statement}
Only facts can be known to be true. If something is known to be true, it is a fact.
\begin{cequation}{a \in \ets}
a \in \Ktrus{i} \then a \in \trus.
\end{cequation}
\end{statement}

This is the notion of propositional knowledge, the most elementary form of knowledge. Propositional knowledge is knowledge of facts, but many of the things that we say that we know are not facts, for example, to know how to do something is not to know a fact; to know a theory is knot to fact. Only propositions can be facts. A theory is not a proposition. 

This theory cannot talk about the other types of knowledge. They are probably in a new layer of complexity, or maybe knowledge is not the best word to say what they are.
\subsection{Acquiring Knowledge}

When a proposition \prp{p} is true, when the observer $i$ finds that \prp{p} is true, its state of knowledge $\state_i$ is updated to $\state|\prp{p} = \prp{p}\state_i\prp{p}$. 

The process of acquiring knowledge about the proposition \prp{p}\ is the transition of one initial possible state of knowledge $\state_i$ to another possible state of knowledge $\state_{i}^{'}$:
\begin{equation}
\state_i \mapsto \state_{i}^{'} = \rst{\state_i{}}{\prp{p}}.
\end{equation}
This process can only occurs if the proposition is true, if not, the new state of knowledge does not describe reality.

\section{Probability}
\begin{formalism}
Any function \pbs\ from propositions to real numbers satisfying:
\begin{enumerate}
\item $\forall \prp{a} \in \prps: \quad \pb{\prp{a}}{} \in \Rset \geq 0$,
\item $\pb{1}{} = 1$,
\item $\forall (\prp{a},\prp{b}) \in \prps\cpt \prps: \quad \pb{\prp{a}\por\prp{b} }{} = \pb{\prp{a}}{} + \pb{\prp{b}}{}- \pb{\prp{a}\pand\prp{b}}{}$,
\end{enumerate}
defines what we call a probability measure.
\end{formalism}

\begin{interpretation}
If \pbs\ is a probability measure and \prp{a}\ is a proposition, then we call \pb{\prp{a}}{} the [probability of \prp{a}].
\end{interpretation}

\begin{interpretation}
We call \pb{\prp{a}}{\state} the [probability of \prp{a} according to $\state$].
\end{interpretation}
\begin{statement}
The probability of a proposition is a non-negative real number. The probability of a proposition is smaller than or equal to one.
\begin{gather}
\pb{\prps}{} \subset \Rset,\\
\forall \prp{a} \in \prps: \quad \pb{\prp{a}}{} \leq 1.
\end{gather}
\end{statement}
\begin{statement}
The probability of the disjunction of two contradictory propositions is the sum of the probabilities of each proposition.
\begin{cequation}{(\prp{a}, \prp{b}) \in \prps \ctrd \prps}
\pb{\prp{a}\por\prp{b} }{} = \pb{\prp{a}}{} + \pb{\prp{b} }{}.
\end{cequation}
\end{statement}

\begin{statement}
The probability of the negation of a proposition is one minus the probability of the proposition.
\begin{cequation}{\prp{a} \in \prps}
\pb{\pnot\prp{a} }{} = 1-\pb{\prp{a}}{} .
\end{cequation}
\end{statement}

\begin{statement}
If \prp{a}\ affirms \prp{b}\ then the probability of \prp{a}\ is smaller than or equal to the probability of \prp{b}.
\begin{cequation}{\prp{a}, \prp{b} \in \prps}
\prp{a}\pleq \prp{b} \then \pb{\prp{a}}{} \leq \pb{\prp{b}}{}.
\end{cequation}
\end{statement}

\begin{formalism}
We define the general probability measure $\pbs_\state$ as a function of a positive operator \state\ by 
\begin{cequation}{\state \in \pops\bsp \prp{a} \in \prps}
\pb{\prp{a}}{\state} = \frac{\msr{\state \prp{a}}}{\msr{\state}}.
\end{cequation}
\end{formalism}

\begin{interpretation}
The [probability measure generated by \state]is denoted by $\pbs_\state$.
\end{interpretation}

For any probability \pbs\ measure one can find a positive operator \state\ such that the probability measure generated by \state\ is equivalent to \pbs.

Any probability measure \pbs\ that satisfies
\[
\pb{\unity}{} =1,
\]
where \unity\ is a projector, is what we call a probability measure of type \unity.

\begin{statement}
Any probability measure generates by a positive operator characterized by \unity\ is a probability measure of type \unity.
\begin{cequation}{\state \in \poprt}
\pbr{\unity} = 1.
\end{cequation}
\end{statement}

\begin{statement}
If \unity\ affirms \prp{a}, then the probability of \prp{a} is one in every probability measure of type one.
\begin{cequation}{\unity \in \pjs\bsp \prp{a} \in \prps}
\state \in \poprt \tand \unity \pleq \prp{a} \then \pbr{\prp{a}} = 1.
\end{cequation}
\end{statement}

\begin{statement}
If \unity\ negates \prp{a}, then the probability of \prp{a} is zero in every probability measure of type one.
\begin{cequation}{\unity \in \pjs\bsp \prp{a} \in \prps}
\state \in \poprt \tand \unity \pleq \pnot\prp{a} \then \pbr{\prp{a}} = 0.
\end{cequation}
\end{statement}

\subsection{Classical Probability}

\begin{interpretation}
When a proposition \prp{P} is compatible with $\state$, we say that [$\pb{\prp{a}}{\state}$ is a classical probability], or that the [probability of \prp{A}\ according to $\state$ is classical].
\end{interpretation}

Classical probability theory arises when we restrict ourselves to a set of compatible propositions with classical probabilities which constitutes a boolean algebra. Now we establish the connection between this formalism and classical probability theory. We will use the same names are used in the literature, however this is just for helping understanding.

\begin{formalism}
Let $\pbs_\state$ be a probability measure of type \unity, then \unity\ is what we call the sample space. 
Let \Xset\ be a boolean algebra contained in \pjsrt and formed by propositions with classical probability, then \Xset\ is what we call the event space. A member of \Xset\ is what is called an event.
\begin{gather}
\state \in \poprt\\
\Xset \pand \Xset = \Xset\dsp\Xset \por \Xset = \Xset\dsp \pcmp{\unity}\Xset = \Xset.
\end{gather}
\vspace{-1.5\baselineskip}
\begin{cequation}{\prp{A} \in \Xset}
\prp{A} \cpt \state
\end{cequation}

Then the triple \tuple{\unity,\Xset,\pbs_\state}\ is a probability space with sample space \unity, event space \Xset\ and probability measure $\pbs_\state$.
\end{formalism}

This structure contains precisely the algebraic structure of classical probability theory. However, Kolmogorov's original formulation\cite{Kolmogorov56} events are sets, but since sets posses the algebraic structure of a boolean algebra, one may find a set of sets that is isomorphic  to \Xset, therefore, for every proposition of \Xset\ there is an isomorphic
set. When this isomorphism is performed, the operation of conjunction and disjunction becomes respectively intersection and union; and the rank of the projector becomes the number of elements in the set. 

Curiously, Kolmogorov uses the notations $AB$ for the conjunction, $A+B$ for the disjunction when $AB=0$ and $\unity - A$ for the complement, thus, for most practical means, there is no difference in both formulations. Actually, classical probability theory is more elegant in this formalism than using sets for representing events, because in this case, the notation used by Kolmogorv becomes really an algebra.

Therefor, rigorously, classical probability theory arises as a particular case of the presented theory. It is a derived theory.

\subsection{The Fundamental Probabilities}

Any probability associated with an object must be of the type of the object.

There is no unique interpretation of probability. It is just a class of mathematical functions with no defined meaning. The meaning of it depends on how it is defined, where it is applied and possibly other factors.

Given an universe there are three fundamental classes of probability measures, each one with a distinct special meaning. Each one of them has a special name that tries to unveil a bit of its meaning. They are:
\begin{enumerate}
\item Possibility: \pbns
\item Plausibility: \pbkis
\item Potentiality: \pbts
\end{enumerate}

Each one of them is a probability measure of the type of the universe generate by a positive operator with a distinct meaning: the type of the universe, a state of knowledge and reality.

\subsection{Possibility \pbns}

The possibility measure \pbns\ is define as the probability measure generated by the type of the universe:
\begin{equation}
\pbns \eqd \pbs_\unity.
\end{equation}
It give us the degree of possibility of a proposition, the possibility of a proposition is one when it is necessary and zero when it is impossible.
When it is greater than zero, the proposition is possible.

The possibility of a proposition \prp{a} is then given by
\begin{cequation}{\prp{a} \in \prps}
\pbn{\prp{a}} = \frac{\msr{\unity \prp{a}}}{\msr{\unity}}.
\end{cequation}

\begin{interpretation}
Given a proposition \prp{a}, we say that \pbn{\prp{a}}\ is [the possibility of \prp{a}].
\end{interpretation}

\begin{statement}
If the possibility of a proposition is one then it is a necessary truth. If the possibility of a proposition is zero then it is a necessary falsity.
\begin{cequation}{\prp{a} \in \prps}
\begin{aligned}
\pbn{\prp{a}}=1 &\iff \prp{a} \in \Ntrus, \\
\pbn{\prp{a}}=0 &\iff \prp{a} \in \Nfalss.
\end{aligned}
\end{cequation}
\end{statement}

\subsection{Plausibility \pbkis}

The plausibility measure \pbkis\ is defined for each observer $i$ as the probability measure generated by its state of knowledge $\state_i$:
\begin{equation}
\pbkis \eqd \pbs_{\state_i}.
\end{equation}

It give us the degree of certainty or how plausible a proposition is, since the proposition is known to be true when the probability is one and known to be false when the probability is zero.

The plausibility of a proposition \prp{a}  according to an observer $i$ is then given by:
\begin{cequation}{\prp{a} \in \prps}
\pbki{\prp{a}} = \frac{\msr{\state_i \prp{a}}}{\msr{\state_i}}.
\end{cequation}

\begin{interpretation}
Given a proposition \prp{a}, we say that \pbki{\prp{a}}\ is [the plausibility of \prp{a} according to the observer $i$].
\end{interpretation}

\begin{statement}
If the plausibility of a proposition is one then it is a known truth. If the possibility of a proposition is zero then it is a known falsity.
\begin{cequation}{\prp{a} \in \prps}
\begin{aligned}
\pbki{\prp{a}}=1 &\iff \prp{a} \in \Ktrus{i}, \\
\pbki{\prp{a}}=0 &\iff \prp{a} \in \Kfalss{i}.
\end{aligned}
\end{cequation}
\end{statement}

\subsection{Potentiality \pbts}

The potentiality measure \pbts\ is defined as the probability measure generated by reality:
\begin{equation}
\pbts \eqd \pbs_{\rlty}.
\end{equation}

It give us the degree of truth of a proposition, since the proposition is true when the probability is one, and is false when the probability is zero.

The potentiality of a proposition \prp{a}\ is then given by:
\begin{cequation}{\prp{a} \in \prps}
\pbt{\prp{a}} = \msr{\rlty \prp{a}} = \bra{\psi} \prp{a}\ket{\psi}.
\end{cequation}

\begin{interpretation}
Given a proposition \prp{a}, we say that \pbt{\prp{a}}\ is [the potentiality of \prp{a}].
\end{interpretation}

In some sense it relates to the idea that the proposition may have a potential of becoming true after a measurement is performed.

This is the probability measure that is used when we are talking about wave functions and state vectors, and is the one that is present on most of time in quantum mechanics.

\begin{statement}
If the plausibility of a proposition is one then it is a truth. If the possibility of a proposition is zero then it is a falsity.

\begin{cequation}{\prp{a} \in \prps}
\begin{aligned}
\pbt{\prp{a}}=1 &\iff \prp{a} \in \trus, \\
\pbt{\prp{a}}=0 &\iff \prp{a} \in \falss.
\end{aligned}
\end{cequation}
\end{statement}

The classical propositions are precisely those whose potentiality is either zero or one, a notion that people usually refer as determinism. I believe this term dos not apply here.

\begin{statement}
The potentiality of a classical proposition is either zero or one.
\begin{equation}
\pbt{\Cprps} = \bools.
\end{equation}
\end{statement}
\begin{statement}
The possibility, the plausibilities and the potentiality of a necessary truth are always one.
\begin{equation}
\pbn{\Ntrus} = \tauts\dsp \pbki{\Ntrus} = \tauts\dsp \pbt{\Ntrus} = \tauts.
\end{equation}
\end{statement}
\begin{statement}
The possibility, the plausibilities and the potentiality of a necessary falsity are always zero.
\begin{equation}
\pbn{\Nfalss} = \ctrs\dsp \pbki{\Nfalss} = \ctrs\dsp\pbt{\Nfalss} = \ctrs.
\end{equation}
\end{statement}
\begin{statement}
The the plausibility and the potentiality of a known truth are always one.
\begin{equation}
 \pbki{\Ktrus{i}} = \tauts\dsp\pbt{\Ktrus{i}} = \tauts.
\end{equation}
\end{statement}
\begin{statement}
The  plausibility and the potentiality of a known falsity are always zero.
\begin{equation}
\pbki{\Kfalss{i}} = \ctrs\dsp\pbt{\Kfalss{i}} = \ctrs.
\end{equation}
\end{statement}
\begin{statement}
The potentiality of a necessary truth is always one.
\begin{equation}
\pbt{\trus} = \tauts.
\end{equation}
\end{statement}
\begin{statement}
The potentiality of a necessary falsity is always zero.
\begin{equation}
\pbt{\falss} = \ctrs.
\end{equation}
\end{statement}

\chapter{Discussions and Conclusions}

The theory presented here represents a paradigmatic shift on what means to interpret and formulate a closed physical theory. It provides us a new interpretation of (part of) quantum theory that is entirely different in nature from any previous interpretation. In other words, the way the interpretation was formulated is entirely different. We claim that all the language used to talk about a theory is part of its interpretation; and that all we can properly say about a theory follows from its interpretation.

It explains why quantum theory (QT) is considered so deep; and why its mathematical formalism is so elegant. Part of quantum theory belongs to the final or fundamental theory of physics (FTP), and this is the most fundamental theory possible in physics. There is no simpler theory. The removal of a single axiom may destroy the whole theory. Most of the interpretation of QT is actually an interpretation of the FTP, and so, it is deeper than any other theory.  \parbreak

The proposed theory provides the basis for a new logic for quantum theory, a logic that is not present in previous formulations, neither in quantum mechanics, neither in the classical logic. The logic proposed here can deal with truth and falsity, necessity and possibility, and knowledge in a unified way. Actually, this would mean kind of a unification of propositional logic, modal logic and epistemic logic, but currently these logics are treated in completely different contexts. Therefore, our approach is different from any previous approach to quantum logic and it incorporates new elements.\parbreak

The Schrödinger's cat model is a good example where we can apply some of the ideas presented here. Actually we are just talking about the superposition of ``dead'' and ``alive'' states with no interest in the entanglement of the cat and any other thing, in fact, we are considering that the system is not entangled. The Hilbert space that describes this system is generated by the orthogonal basis \sdef{\ket{D}, \ket{A}}. In our language, the cat is an universe (i.e. not entangled), with the type $\unity = \stp{D} + \stp{A}$. Let's suppose that the vector state that describes the cat is given by $\ket{\psi} = \frac{1}{\sqrt{2}}(\ket{D} + \ket{A})$. In our language, it means that the reality of this universe, or, the state of the cat, is given by $\rlty = \frac{1}{2}(\stp{D} + \ket{D}\bra{A} + \ket{A}\bra{D} + \stp{A})$.

The question an interpretation should answer is: what can we say about this system? Can we say that the cat is alive? Can we say that it is dead? That it is dead or alive? Or all we can say is the probability of opening the box and finding the cat dead or finding it alive? In this case, what can we say after we opened the box? Say once more the probability of what will occur if we open the box again? Note that probability here cannot be related to uncertainty since the state of the system was already given, and has zero entropy.

According to  Copenhagen interpretation, all we can say are the probabilities of each result of an experiment. But this is to minimalist. What can we say before any experiment is performed, given the state of the system?

Using the interpretation presented here, we can understand the situation once we define mathematically things we were just saying. That is, what means to say that the cat is dead? It means that the proposition associated to this sentence is true. Therefore, we define the propositions:
\begin{align}
\text{``the cat is dead''} &= \stp{D}\\
\text{``the cat is alive''} &= \stp{A}
\end{align}
And also we have
\begin{align}
\text{``the cat is dead or alive''} &= \stp{D}\por\stp{A} = \unity\\
\text{``the cat is dead and alive''} &= \stp{D}\pand\stp{A} = 0
\end{align}

Now, we can say that the cat is dead if, and only if the proposition ``the cat is dead'' is true, and the same for the other sentences. From the definition of the state, we can show that: $\stp{D} \npgeq \rlty$, $\stp{A} \npgeq \rlty$, $\unity \pgeq \rlty$ and $0 \npgeq \rlty$.  Using the definition of truth and falsity we have that:
\vspace{-\parskip}
\begin{enumerate}
\item ``the cat is dead'' is not true and not false,
\item ``the cat is alive'' is not true and not false,
\item ``the cat is dead or alive'' is true,
\item ``the cat is dead and alive'' is false,
\end{enumerate}
\vspace{-\baselineskip}
and also that:%
\vspace{-\baselineskip}
\begin{enumerate}%
\item ``the cat is dead or alive'' is necessarily true,
\item ``the cat is dead and alive'' is necessarily false.
\end{enumerate}

Note that we have propositions that are neither true, neither false; and that is something new, because according to classical logic, a proposition is either true or false. Quantum theory violates one of the main principles of classical logic, the \emph{principle of excluded middle.} The theory force us to reconsider our \emph{logic!} However, this is not in the same sense as proposed by the original quantum logic \cite{BkN}. The principles of logic should be changed, not the distributive law. These notions may also provide a final answer to Putnam question ``Is logic empirical?''\cite{Putnam}.  Logic follows from the mathematical formalism of the FTP, and classical logic is not the `right' logic. It is used for understanding and formulating empirical theories, but it is part of the framework, not an empirical theory itself. However, these ideas must be stated in mathematical terms, just then we have a final answer.

The principle of excluded middle is valid precisely for what we defined as classical propositions, and so, if we restrict our domain of propositions to the classical ones, then the classical logic becomes valid and ``we are in a classical world''. But we don't need a different theory to talk about the classical world; the classical and the quantum are not from different natures. One is just a particular case of the other.

Classical propositions are also the only ones we can verify their truth or falsity experimentally. In other words, a yes-or-no experiment can reveal the truth or falsity only of classical proposition, because, in the case of a non-classical proposition, it will necessarily change or disturb the state of the system. In this case we cannot say that the experiment or measurement acted like a true observation just revealing something that was there, or being more precise, revealing something that was true.

In the same way we defined a classical proposition we define a \emph{classical observable} as an observable that is compatible with reality. Then it follows that the classical observables are precisely the ones that \emph{have a defined value.} But what it means for an observable to have a defined value? For an observable $A$ defined as
\[
A = \sum_i a_i \prp{P}_i, \quad \prp{P}_i \prp{P}_j = \delta_{i,j} \prp{P}_i,
\]
we say that the observable $A$ has the value $a_i$ iff $\prp{P}_i$ is true, in other words,
\[
\text{``the observable $A$ has the value $a_i$''} = \prp{P}_i
\]
 Therefore, an observable has a defined value iff one of its projectors is true, a condition that is satisfied if, and only if, the observable is classical. Further results will show that special quantum correlation, like violations of  Bell's inequalities\cite{EPR}\cite{Bell64}\cite{CHSH69}  can only occur between non-classical observables, since between classical observables all the equation will look like the corresponding classical ones.\parbreak

According to the presented theory, there is not a single probability measure associated with a physical system, but at least three kinds of probabilities. One is associated with the identity of system's Hilbert space. One is associated with a state of knowledge of some observer, which is a density operator. And one associated with the state of the system itself.

Each of those probability measures has a distinct meaning, and so there is not an unique interpretation of probability.  We call the probability measure generated by the state of the system, $\pbt{\prp{A}} = \bra{\psi}\prp{A}\ket{\psi}$, as potentiality, since what it measures is something like the potential of a proposition becoming true when a measurement is performed. This one is the most used in quantum mechanics since is the one associated with the so called wavefunction.

Given the status of the presented theory, one cannot consider (classical) probability theory\cite{Kolmogorov56} as a theory that is more fundamental than QT, but just a theory that is part of it. Therefore, probability theory is not something can serve as a foundation for QT, but is QT which gives rise to classical probability theory as a particular case of it. On the other side, the proposed theory is a (general) probability theory, in other words, probability theory is in the essence of the FTP. This is not new since quantum theory can be seen as a probability theory\cite{Pitowsky}, but given the depth of the FTP, this might explain why probability is used everywhere in science.

The proposed theory contains elements of both a \emph{subjective or epistemic} nature (the states of knowledge), and an \emph{objective or ontic} nature (the state of the system). Therefore, it provides answers to the ongoing debate\cite{nn} on the nature of the quantum state, and our claim is that this notion, (quantum state) is ill defined. The importance of the distinction between pure and mixed states, and the fact that only pure states can be associated with a vector (ket) is dismissed. The distinction show totally different mathematical properties and meaning of each of them. We say that the pure state \rlty\ is something objective, actually that it represents reality itself. And the mixed states \state\ represents something epistemic, that they are the state of knowledge of some observers. This two elements are not, however, completely independent, there is a condition for a density operator to be considered a possible state of knowledge. However, the state of knowledge is not totally defined by reality, and that's why we say that it is subjective. We can say that this theory is a theory of knowledge and truth, and this the explanation of why there are elements of this two natures: epistemic and ontic. These notions unscramble the omellete that was mentioned by Jaynes:
\begin{quote}
``But our present [quantum mechanical] formalism is not purely epistemological; it is a peculiar mixture describing in part realities of Nature, in part incomplete human information about Nature -- all scrambled up by Heisenberg and Bohr into an omelette that nobody has seen how to unscramble. Yet we think that the unscrambling is a prerequisite for any further advance in basic physical theory. For, if we cannot separate the subjective and objective aspects of the formalism, we cannot know what we are talking about; it is just that simple.''\cite{Jaynes90}
\end{quote}

It is also compatible with all the recent publications\cite{PBR}\cite{HarriganSpekkens10}\cite{Leifer11}\cite{Hardy12}\cite{Montina08}\cite{Spekkens07}\cite{Colbeck12} in which the quantum state $\ket{\psi}$ is said to represent something real, and again, we say that \rlty\ represents reality itself and that there is no deeper theory that will say what reality or represent it. Since this part of quantum theory is part of the FTP, \emph{if quantum theory does not represent reality, no other theory does!} This is the great deal of considering part of it a final theory; you cannot left the most fundamental questions to be answered by an unknown deeper theory. There are question that will be only answered when we extend this theory, but there is no deeper theory because there is no simpler theory. We have reached a limit.

In the language of some recent publications\cite{HarriganSpekkens10}\cite{Hardy12}, the proposed theory is a kind of $\psi$-ontic and $\state$-epistemic model, and therefore it has the benefits of both the epistemic and ontic interpretations of the quantum state, since we say that there are actually two kinds of quantum state: state of the system and state of knowledge. The collapse occurs in the state of knowledge when the observer finds that another proposition is true. In this case no change occurs in the system's state, and the collapse means simply that the observer acquired knowledge. There is no mysterious physical process. What happens is something we may call an epistemic processes. However, the measurement process is an interaction with the system, and when we try to measure a non-classical observable, this interaction necessarily changes the state of the system.

A measurement always give a result that at least appears to be a defined value, however, a non-classical observable has not a defined value. The idea that every physical quantity has a defined value is a strictly classical idea, but an idea that is part of our worldview. And this notion follows directly from believe in the principle of excluded middle. It is our worldview what introduces so many paradoxes in quantum theory.\parbreak

Assuming that the proposed theory is really a final theory, in a sense, it rules out any attempt of formulating a hidden variable model consistent with the results of QM. What occurs is that, \emph{any mathematical theory that reproduces the results of QM is either QM itself, or a theory that is not as simple and elegant.} By the Ockham's razor principle or the ideal of beauty, we should prefer quantum mechanics than any alternative. In other words, \emph{there is no other mathematical theory capable of reproducing all the results of QM that is more beautiful or simpler than quantum mechanics.}

However, the ideal of beauty can only be applied in an objective way when we have theories in their closed formulation, if not, we might never be sure about which theory is simpler. When a physical theory is formulated in its closed form, the theory is about to reach its full potential in terms of conceptual clarity and mathematical rigor; and this is how we should try to formulate our best physical theories. The proposed theory is probably the first one in a closed form, or at least very close to it. However, it contains only a fraction of today's QT. Classical mechanics can also be formulated in a closed form using the proposed framework\cite{Koopman}\cite{Gozzi}. And once we have the closed formulations of classical and quantum mechanics, we will be able to say precisely where they differ, which axioms are different and which are not. \emph{The difference between classical and classical mechanics will be stated in mathematical terms,} with no room for doubts.\parbreak

This work proposes not only a new interpretation but a new approach for interpreting physical theories. In this approach, most of the philosophical preconceptions are avoided, and the main objective of the interpretation is to answer the questions: what can we say about the theory? What can we say about each of its mathematical statements?\parbreak

We have provided a new closed formulation of part of quantum theory, that is, we gave a completely axiomatic formulation of all the mathematical formalism of the theory and provided a way of interpreting each of its mathematical statements. In the formalism, Dirac's notation is no longer simply a notation, but an algebra properly. The C$^*$-algebra formalism, and the usual bra - ket formalism are completely unified. All the interpretation is simply a reading of its mathematical formalism. There is no room for ambiguities. And that is why we claim that this is a final theory, that there is no way of making it more fundamental.

According to this approach, our closed theories are not supposed to be based on concepts that have not been mathematically defined, they should define or give meaning to the fundamental concepts used in their interpretations. Since the proposed theory gives meaning to fundamental concepts, we claim that it is the basis for a new worldview, \emph{the worldview that makes quantum theory understandable.}\parbreak

The proposed theory has no underlying ontology and thus, makes no ontological commitment. Ontology is still not a well-defined philosophical notion, and it cannot provide any basis for the FTP. Actually, the proposed theory is the basis for a theory of ontology. It might help us find  a clear, and precise meaning to ``existence'' and ``being''.

The distinct concepts we called Categories are exactly some of the categories of being which are studied by ontology, but now we gave them  mathematical significance. Categories are concepts that are independent on how our world is; which is different from Truth, a concept that depends on how reality is. The theory defines some categories in which things can exist, but it says nothing about what actually exists. This is an empirical question. Science or our own experience should say what really exists. Ontology can only tell us how to understand or classify the things that exist or could exist. \parbreak

From the development of this notion of objective interpretation, a new theory on formal interpretation of closed theories may arise. This theory will help us to understand the underlying connection between math and ordinary language. Once we have it, we will be able to give  formal semantics to part of our natural languages. The interesting point is that, the languages we use to talk about our closed theories, and so, the language of physics, will be the first ones we can give a formal semantics. Then math gives meaning to ordinary language, and ordinary language gives meaning to math! Interpreting is just the translation from one language to the other! The interpretation of the fundamental physical theories can even provide  formal semantics for ordinary language!\parbreak

Within the proposed framework, we could mathematically give  meaning to the word ``reality''. The connection of reality with all the other fundamental concepts defined here forms the nature of reality. What we can say in plain words, is that reality is the actual state of the universe. To understand the nature of reality is not to have knowledge about it, but to go beyond a simple word and dig into its true meaning. That is understanding. Experience is the process of knowing how Reality is. This is an endless activity. It is not possible for anyone to experience everything or know every fact. That's why we created Science: to compress facts and minimize the need for experience. With Science, each new generation has no longer need to experience everything the last generation has experienced. It can continue where the last one stopped.
\begin{center}
$\infty$
\end{center}


%
%


\begin{thebibliography}{99}


\bibitem{Deutsch98} Deutsch, David, ``The Fabric of Reality: The Science of Parallel Universes--and Its Implications'', (Penguin, 1998).



\bibitem{Weiz} Weizsäcker, Carl F. von, ``The Structure of Physics'', Fundamental Theories of Physics, Vol. 155 (Springer, 2006)

\bibitem{Feynman65} Feynman, Richard. ``The Character of Physical Law'', (1965)

\bibitem{Weinberg94} Steven Weinberg,  ``Dreams of a Final Theory: The Scientist's Search for the Ultimate Laws of Nature'' , (Vintage, 1994)

\bibitem{Fuchs02} Fuchs, Christopher A.,  ``Quantum Mechanics as Quantum Information (and only a little more)'', \texttt{ quant-ph/0205039v1}, (2002).

\bibitem{Heisenberg72} Heisenberg, W., ``The Correctness-Criteria for Closed Theories in Physics'', In W. Heisenberg \emph{Encounters with Einstein: And Other Essays on People, Places, and Particles.}  Princeton, NJ: Princeton University Press, (1972).

\bibitem{Heisenberg58} Heisenberg, W., ``Physics and Philosophy'', (Penguin, 2000)


\bibitem{Kuhn} Kuhn, Thomas S.  ``The Structure of Scientific Revolutions'', International Encyclopedia of Unified Science, (1962)

\bibitem{Feynman82} Feynman, Richard. ``Simulating Physics with Computers'',  International Journal of Theoretical Physics (1982).


\bibitem{Kolmogorov56} Kolmogorov, Andrey. ``Foundations of the Theory of Probability'' (2nd ed.). (Chelsea, New York, 1956).



 

\bibitem{Neumann32}von Neumann, J. (1932). ``Grundlagen der Quantenmechanik'', Berlin: Springer Verlag (English translation:
``Mathematical foundations of quantum mechanics'', New Jersey: Princeton University Press, 1996).

\bibitem{Cohen} Tannoudji, C. \textbf{Quantum Mechanics Vol. 1}, (1985).




\bibitem{Jaynes90} E. T. Jaynes, in Complexity, Entropy, and the Physics of Information, edited by W. H. Zurek (Addison-Wesley, 1990) p. 381.

\bibitem{Wittgenstein} Ludwig Wittgenstein, ``Tractatus Logico-Philosophicus'', (1922)

\bibitem{BkN} G. Birkhoff and J. von Neumann, ``The Logic of Quantum Mechanics'', Annals of Mathematics Vol. 37, pp. 823--843, 1936.

\bibitem{Putnam}Putnam, H., ``Is Logic Empirical?'', Boston Studies in the Philosophy of Science, vol. 5, eds. Robert S. Cohen and Marx W. Wartofsky (Dordrecht: D. Reidel, 1968).

\bibitem{Bell64} J. S. Bell, ``On the Einstein Podolsky Rosen Paradox'', Physics 1, 3, 195-200 (1964)

\bibitem{EPR} A. Einstein, B. Podolsky, and N. Rosen. ``Can Quantum-Mechanical Description of Physical Reality Be Considered Complete?'', Phys. Rev. 47, 77-780 (1935)

\bibitem{CHSH69}J. F. Clauser, M.A. Horne, A. Shimony and R. A. Holt, ``Proposed experiment to test local hidden-variable theories'', Phys. Rev. Lett. \textbf{23}, 880-884 (1969)

\bibitem{Gozzi} E. Gozzi, D. Mauro, ``On Koopman-von Neumann Waves II'', Int.J.Mod.Phys. A19 (2004) 1475-1494, arXiv:quant-ph/0306029v2

\bibitem{Koopman} B. O. Koopman, Proc. Natl. Acad. Sci. U.S.A. 17, 315 (1931).

\bibitem{PBR} Pusey, M. F., Barrett, J., and Rudolph, T. (2011), On the reality of the quantum state, Nature Phys. \textbf{8},  arXiv:1111.3328.

\bibitem{HarriganSpekkens10} N. Harrigan and R. W. Spekkens, ``Einstein, Incompleteness, and the Epistemic View of Quantum States'', Found. of Phys. \textbf{40},  pp. 125--157, (2010)

\bibitem{Leifer11} M. Leifer, ``PBR, EPR, and all that jazz'', The Quantum Times \textbf{6}, 1 (2011).

\bibitem{Montina08} A. Montina,``Exponential complexity and ontological theories of quantum mechanics'', Phys. Rev. A \textbf{77}, 022104 (2008). arXiv:0711.4770

\bibitem{Hardy12} Lucien Hardy, ``Are quantum states real?'', arXiv:1205.1439

\bibitem{Spekkens07} R.W. Spekkens, ``Evidence for the epistemic view of quantum states: A toy theory'', Physical Review A, 75(3):032110, 2007.

\bibitem{Colbeck12}R. Colbeck and R. Renner, ``Is a systems wave function in one-to-one correspondence with its elements of reality?'' Physical Review Letters, 108(15):150402, 2012.

\bibitem{Cox46}Cox, R. T. ``Probability, frequency, and reasonable expectation'', Am. Jour. Phys., \textbf{14},1-13 (1946).







\bibitem{Pitowsky} Pitowsky, Itamar. ``Quantum mechanics as a theory of probability'', \texttt{quant-ph/0510095v1}, (2005)




\end{thebibliography}


\end{document}